\documentclass[aps,prd,floatfix,%
 reprint,
superscriptaddress,
 amsmath,amssymb,
 aps,
]{revtex4-2}

\usepackage{orcidlink}
\usepackage{graphicx}
\usepackage{dcolumn}
\usepackage{bm}
\usepackage{xcolor} 
\usepackage{multirow}
\usepackage{hyperref}
\usepackage[mathlines]{lineno}

\usepackage{natbib}
\begin{document}

\preprint{APS/123-QED}

\title{A Morphology-Based Cross-Tracer Ratio for Robust BAO Tests}

\author{Paula S. Ferreira\orcidlink{0000-0002-7540-040X}}
\email{ferreira\_paula@zju.edu.cn}
\affiliation{Institute of Astronomy, School of Physics, Zhejiang University, Hangzhou, China}
\affiliation{Center for Cosmology and Computational Astrophysics, Institute for Advanced Study in Physics, Zhejiang University, Hangzhou, China.}

\author{Carlos A. P. Bengaly\orcidlink{0000-0001-5731-3348}}%
\affiliation{Observatório Nacional,\\
 Rua General José Cristino, 77, São Cristóvão, Rio de Janeiro-RJ. CEP 20921-400, Brazil}

\author{Renyue Cen\orcidlink{0000-0001-8531-9536}}%
\email{renyuecen@zju.edu.cn}
\affiliation{Institute of Astronomy, School of Physics, Zhejiang University, Hangzhou, China}%
\affiliation{Center for Cosmology and Computational Astrophysics, Institute for Advanced Study in Physics, Zhejiang University, Hangzhou, China.}

\date{\today}

\begin{abstract}
We use the Cross-Tracer Clustering Ratio (CTCR) acoustic response estimator, a new probe of morphology-dependent galaxy clustering in photometric surveys. CTCR is the ratio of the auto-power spectrum of early-type galaxies (ETGs) to their cross-power spectrum with late-type galaxies (LTGs). In the matched-kernel deterministic linear-bias limit the CTCR cancels the complete projected matter spectrum, including the baryon acoustic oscillation (BAO) wiggles, leaving only the relative linear bias. This makes the ratio uniquely sensitive to physics beyond deterministic linear bias: any residual acoustic structure in the CTCR must arise from a difference between the normalised non-linear responses of the two tracers. We, for the first time, exploit this property to model the measured ratio as a smooth scale-dependent baseline multiplied by a scale-dependent acoustic template, and constrain the amplitude \(A_{\rm morph}\) of the differential acoustic response jointly with the BAO dilation parameter \(\alpha\). Applying the estimator to nine tomographic bins of the DESI Legacy Survey Data Release 8 spanning \(0.1 < z < 0.9\), we obtain \(\alpha = 0.9997 \pm 0.0383\), consistent with the Planck 2018 \(\Lambda\)CDM prediction, and \(A_{\rm morph} = 0.369 \pm 0.307\), consistent with zero. A Bayesian model comparison  yields at most a \(2.6\sigma\) preference for the acoustic template over a smooth-only baseline. On the Euclid Flagship 2 mock, the morphology-selected pair gives \(A_{\rm morph} \simeq 0\) and \(\alpha \simeq 1\) across all redshift bins, while mass-selected pairs with mismatched redshift distributions show a mild negative response. The estimator is model-independent, requires no external calibration of the absolute bias, and is directly applicable to the wide-field imaging surveys that will define the next decade of cosmology. It offers a new route to testing halo occupation and assembly-bias models, and an independent, systematics-robust cross-check of BAO measurements from spectroscopic surveys.
\end{abstract}
\keywords{PACS: 98.80.-k, 98.80.Es, 95.36.+x; Keywords: Cosmology, Baryon Acoustic Oscillations, Galaxy clustering, Cross-correlation, Multi-tracer technique, Dark energy, DESI Legacy Survey}

\maketitle

\section{Introduction}

The precision era of cosmology was built upon the foundational measurements, such as from the Planck satellite \cite{2020}, the Sloan Digital Sky Survey's BOSS and eBOSS \cite{mgs,Raichoor_2020,Alam_2017,Hou_2020} and the Dark Energy Survey (DES) \cite{2026arXiv260114559D}, which together helped stablish the standard cosmological model with great accuracy. This legacy is now being advanced by current experiments such as \textit{Euclid} \cite{euclid}, the Vera C. Rubin Observatory (LSST) \cite{abell2009lsst}, and  Dark Energy Spectroscopic Instrument (DESI) \cite{Abdul_Karim_2025}, and will reach even greater heights with future missions including the China Space Station Telescope (CSST) \cite{csst}, the Nancy Grace Roman Space Telescope \cite{bartusek2022nancy}, and the Square Kilometre Array (SKA) \cite{weltman2020fundamental}. One of the primary scientific objectives of these surveys is to precisely measure the Baryon Acoustic Oscillations (BAO) and thereby constrain the expansion history of the Universe. The BAO is a consequence of decoupling between baryons and photons at $z\simeq 1100$. When no longer coupled and forming acoustic oscillations in the energy density field, the last sound wave-front left was frozen forcing galaxies to cluster close to the spherical shell made out of this wave-front of radius of about 150 Mpc in comoving distance, considering the $\Lambda$ Cold Dark Matter ($\Lambda$CDM) model. 

Thanks to the BAO fixed scale, it is possible to constrain cosmological parameters to describe the universe's expansion history. The statistical tool to find the BAO directly from the perturbation theory derivation is the Power Spectrum, which is defined as the variance of the density fluctuations. Any excess of clustering increases the power spectrum, while absence of clustering suppresses it. 

In three-dimensional Fourier space, the measurement of the power spectrum, $P(k)$, requires a precise distance relation for an accurate measurement, which is well obtained by spectroscopic redshift estimation. The disadvantage of spectroscopic redshift is the amount of exposure time, limiting the number of observed objects, a complication for reaching the desired fair sample of the universe. In contrast, photometric surveys allow for many more targets; for instance, the number of Luminous Red Galaxies (LRGs) increases by a factor of seven when moving from the combined BOSS+eBOSS sample to DES \cite{Rosell_2021}. However, with lower resolution the increase of redshift uncertainty smears the BAO signal in the line-of-sight, the possible solution for this issue, as proposed here using a new technique, is to measure the transverse BAO feature through the angular power spectrum ($C_\ell$) or its Fourier transform the angular correlation function $w(\theta)$.

The application to real-space observations of the angular BAO size ($\theta_{\rm BAO}$) was first performed by \citet{Sanchez_2010} and later by \cite{Crocce_2011,carnero2011clusteringphotometricluminousred}. The results indicated a substantial sensitivity to the Chevallier-Polarski-Linder (CPL) parametrization of the Dark Energy equation of state. In this framework, Dark Energy is described by two parameters, $w_0$ and $w_a$, which govern its time evolution according to $w(a)=w_0 + (1-a)\, w_a$. This formalism enables detailed studies of how the properties of Dark Energy change over cosmological timescales \cite{chevallier2001accelerating,linder2003exploring}. Presently, with a higher resolution, measurements from the DESI DR2, when combined with CMB constraints, concluded that there is a 3.1$\sigma$ preference for CPL over the $\Lambda$CDM model \cite{Abdul_Karim_2025}. Even though the combination of these precise experiments is already worth considering an update to the current model for the universe, thorough analysis with possibly improved or independent probes to check such an important claim is desirable.

The main results of the transverse BAO scale were found by the DES collaboration \cite{DESY6}. Using the BAO shift parameter, $\alpha$, defined as the ratio of the measured angular distance scale to the Planck \cite{2020} $\Lambda$CDM prediction, $\alpha \equiv (D_M/r_d)_{\text{measured}} / (D_M/r_d)_{\text{Planck}}$, such that $\alpha = 1$ corresponds to perfect agreement with the standard model. The DES Y6 analysis measured this parameter using three independent and blinded pipelines: the Angular Correlation Function (ACF), the Angular Power Spectrum (APS), and the Projected Correlation Function (PCF), yielding values of $\alpha_{\rm ACF} = 0.952 \pm 0.023$, $\alpha_{\rm APS} = 0.962 \pm 0.022$, and $\alpha_{\rm PCF} = 0.955 \pm 0.020$, respectively. These individual measurements are highly consistent with each other, providing strong internal cross-validation of the result. The combined consensus from these three methods is $\alpha = 0.957 \pm 0.020$, which corresponds to a measured BAO scale that is $4.3\%$ smaller than the Planck $\Lambda$CDM prediction, a mild discrepancy significant at the $2.1\sigma$ level. 

All the surveys aforementioned prioritized the individual tracers clustering above cross-correlating different tracers. The most obvious reason is that most tracers are found in different redshift distribution or footprint. However, the multi-tracer technique offers several distinct advantages over traditional single-tracer analyses. First and foremost, by combining multiple biased tracers that respond differently to the underlying matter distribution, one can effectively cancel sample variance.  This cancellation arises because the cross-correlations between different populations contain additional information about the shared large-scale structure that is inaccessible from auto-correlations alone \cite{Alarcon}. 

\citet{MS} proposed that by observing multiple tracers of large-scale density with different biases (e.g., different galaxy populations that cluster differently), one can evade this limit. Each tracer samples the same underlying dark matter field but with a different bias factor. By comparing their clustering patterns, one can separate the common, cosmic variance signal from the tracer-specific information.

The improvement scales with the bias difference between tracers: in the noiseless limit, multiple tracers can greatly improve constraints on the dark energy equation of state and the growth parameter, with the gain increasing rapidly as the bias contrast grows. In realistic survey scenarios, \citet{Alarcon} demonstrated improvements in the Figure of Merit by factors ranging from 1.3 to 9.0, depending on the global galaxy density and the effectiveness of the tracer split. 

Moreover, multi-tracer analyses are particularly powerful when combined with redshift tomography. By splitting a galaxy sample into narrow redshift bins and analysing both auto- and cross-correlations between them, one can recover radial information that is otherwise lost in purely angular analyses. \citet{Loureiro} demonstrated this by analysing the angular power spectra of BOSS DR12 galaxies in 13 tomographic bins, obtaining competitive cosmological constraints, including a $\sim$4\% measurement of the dark energy equation of state. The cross-correlation of different tomographic bins provides additional leverage on the expansion history, as the overlap between bins encodes information about the radial selection function. 

By combining observables that trace the large-scale structure in complementary ways, one can break degeneracies and significantly improve parameter constraints. \citet{arf} proposed the combination of the standard Angular Density Field (ADF) with the recently proposed Angular Redshift Fluctuations (ARF). It measures the angular fluctuations in the galaxy redshift field, providing a probe that is sensitive to both the underlying matter density and the peculiar velocity field while remaining largely uncorrelated with the standard galaxy angular density field. For narrow redshift windows, the ARF field is found to be almost completely uncorrelated with the ADF, meaning that they carry independent information about the underlying density and velocity fields. This complementarity translates into a substantial gain in constraining power: a joint ADF+ARF analysis increases the constraints by at least an order of magnitude for parameters such as the Hubble constant \(H_0\) and the dark energy equation of state parameters \(w_0\) and \(w_a\) \cite{arfbao}. Crucially, to fully exploit this information, it is necessary to include cross-correlations between neighbouring redshift shells, with the information converging only when correlations extending up to \(\Delta z \sim 0.6\) for ADF and \(\Delta z \sim 0.4\) for ARF. 

In this work, we explore the advantages of cross-correlating different tracers of the large-scale structure through the Cross-Tracer Clustering Ratio (CTCR) \cite{Ferreira:2026iru}, defined as the ratio of the auto-power spectrum of tracer \(i\) to its cross-power spectrum with tracer \(j\),$\mathcal{R}_\ell^{ij}     \equiv    \frac{C_\ell^{ii}}{C_\ell^{ij}}$. In the matched-kernel deterministic linear-bias limit, the CTCR cancels the complete projected matter spectrum, including both the smooth broadband shape and the BAO wiggles, leaving only the ratio of the two linear biases. Any residual scale dependence in the ratio must therefore originate from physics beyond linear bias, such as a difference between the normalised non-linear responses of the two tracers. We exploit this property to construct an \emph{acoustic response estimator}: we model the measured ratio as a smooth scale-dependent baseline multiplied by a scale-dependent acoustic modulation, and constrain the amplitude \(A_{\rm morph}\) of the differential acoustic response together with the BAO dilation parameter \(\alpha\). We apply the estimator to early-type (ETG) and late-type (LTG) galaxies in the DESI Legacy Survey Data Release 8 \citep{desi_legacy}, using nine tomographic bins spanning \(0.1 < z < 0.9\), and validate the pipeline on the Euclid Flagship 2 mock\cite{Euclid:2024few}. The resulting constraint on \(\alpha\) is consistent with the Planck 2018 \(\Lambda\)CDM cosmology, and \(A_{\rm morph}\) is consistent with zero, indicating that the two morphological populations do not exhibit a detectable differential
acoustic response at the current level of precision.

\section{Data}
We used the DESI Legacy Survey Data Release 8 \cite{desi_legacy}, together with photometric redshifts from \citet{zhou_DL}. To assign weights to the best objects, we adopted different parameter sets for each morphological type. For spiral galaxies, we used \texttt{rchisq\_r}, \texttt{rchisq\_i}, \texttt{shapeexp\_e1}, and \texttt{shapeexp\_e2}; for elliptical galaxies, we used \texttt{rchisq\_r}, \texttt{rchisq\_i}, \texttt{shapedev\_e1}, and \texttt{shapedev\_e2}.

\begin{figure}
    \centering
    \includegraphics[width=\linewidth]{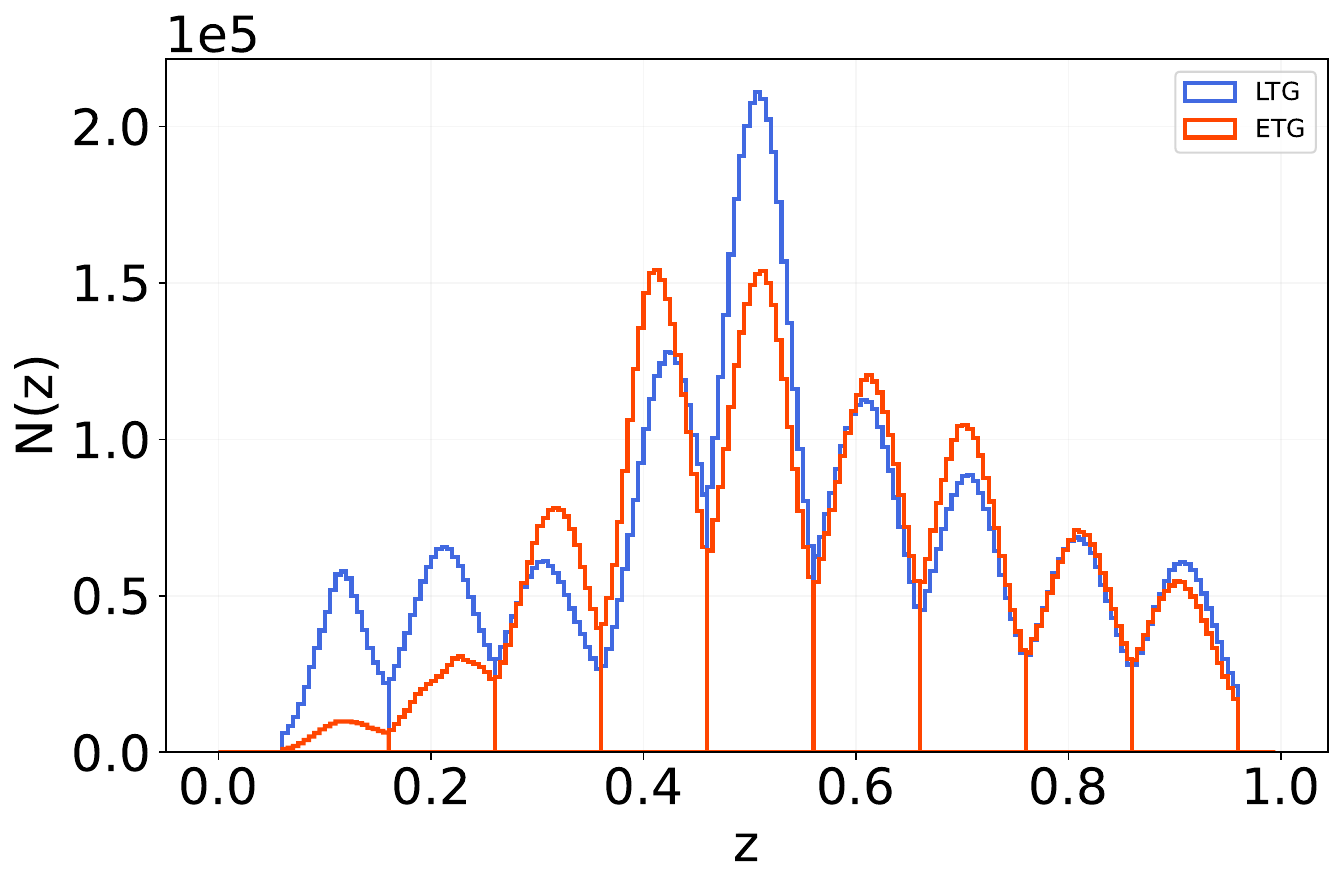}
    \caption{Bin separation of the DESI Legacy DR8 data. The ETG galaxies distribution is shown in red, while the LTG are in blue.}
    \label{fig:bins}
\end{figure}

The random catalog employed for null-signal tests in LSS is generated from a mask specified separately for each redshift bin (see Fig.~\ref{fig:bins}), using \texttt{NSIDE} = 512. This random sample includes 50 times as many objects as the actual data set, with galaxies placed uniformly inside the mask while matching the original sample’s redshift distribution.

To improve the purity of our morphological samples and to reduce the impact of galaxies whose classification is uncertain or corrupted by systematics, we assign a per‑galaxy weight \(w_g\). The weight quantifies how well a galaxy’s observed properties match the expected behaviour of a securely classified object of its morphological type (elliptical or spiral). High‑weight galaxies are considered reliable; low‑weight galaxies are down‑weighted in the clustering analysis.

We define the weight using a linear regression that predicts the normalized redshift \(Z_g = z_g / \bar{z}_g\), where \(z_g\) is the photometric redshift of each galaxy and \(\bar{z}_g\) is the iteratively estimated mean redshift of the tomographic bin. The input features are \(\mathbf{X}_g = \{ w_z,\, x_p \}\), with \(w_z = 1/\sigma_z\) (the inverse photometric‑redshift uncertainty) and \(x_p\) denoting the set of morphological and observational systematics parameters. We then fit a linear model of the form $Z_g = w_g \cdot (\text{linear combination of predictors}) + \text{error}$,
and determine the single coefficient \(w_g\) that minimizes the sum of squared residuals. This best‑fit \(w_g\) is adopted as the weight for that galaxy. Objects with well‑behaved properties (small photo‑z uncertainties, high morphological classification probability, and typical magnitudes—naturally obtain \(w_g \approx 1\), while galaxies with outlying characteristics or poor shape fits are assigned weights close to 0. Although this weighting procedure leaves the overall redshift distribution essentially unchanged, it enhances the morphological purity of the samples, which is crucial for constraining the scale‑dependent bias ratio. The weight \(w_g\) directly modifies the galaxy counts in each pixel: well‑measured galaxies contribute more strongly to the density contrast and therefore to the resulting \(C_\ell\), whereas galaxies with uncertain properties are down-weighted. A detailed validation of these weights is presented in \cite{Ferreira:2026iru}.

The effective mean redshift of each bin is computed iteratively as the weighted average of the individual photo‑\(z\):
\begin{equation}
    \bar{z} \pm \sigma_z= \frac{\sum_g z_g \, w_g}{\sum_g w_g}\pm\frac{\sum_g (z_g^{err}/(1+z_g))\,w_g}{\sum_g w_g}.
\end{equation}
This ensures that the reference redshift for each tomographic bin is consistently defined after the weight application. Our weighted average of $z_{\mathrm{err}}/(1+z_{\mathrm{phot}})$ yields $\sigma_z \approx 0.03$--$0.04$ depending on the redshift bin for the DESI samples, which is larger than \cite{zhou_DL}'s $\sigma_{\mathrm{NMAD}} = 0.02$ for LRGs (based on spectroscopic redshifts), indicating that our approach provides a conservative estimate of the redshift uncertainty in the absence of a spectroscopic calibration sample.

\section{The Angular Power Spectrum estimator}
We estimate angular power spectra using the pseudo-$C_\ell$ method implemented in \texttt{NaMaster} \cite{Alonso_2019} (\texttt{pymaster}). All maps are built at \texttt{NSIDE}=$256$, and spectra are measured over $0<\ell<800$, though only multipoles $20<\ell<200$ are used in the cosmological analysis to stay within the range where the Limber approximation is reliable.

The uncertainties are estimated from $100$ patch‑based bootstrap realisations. For each realisation we randomly select $80\%$ of the HEALPix pixels at \texttt{NSIDE}=$512$ (pixel size $\sim 0.1^\circ$) with replacement, keeping all galaxies in the chosen pixels. This preserves large‑scale correlations better than a galaxy‑by‑galaxy resampling. The resulting empirical covariance is regularised by linear shrinkage \cite{Pope2008,Euclid2025} towards a diagonal target $\mathbf{T}=\operatorname{diag}(\mathbf{C}_{\text{boot}})$ with a fixed shrinkage intensity $\alpha=0.2$, and a small constant term $\epsilon\mathbf{I}$ is added to guarantee invertibility.

 For a given galaxy sample and a corresponding random catalogue (weighted to follow the survey mask and radial selection), we construct the weighted galaxy count map $n_{\rm g}(p)$ and the normalised random map $\bar{n}_{\rm r}(p)$. The overdensity map $\delta(p) = m(p)[n_{\rm g}(p)/\bar{n}_{\rm r}(p)-1]$ is then transformed to spherical harmonics. The resulting pseudo‑$C_\ell$ is related to the true $C_\ell$ by the mode‑coupling matrix $\mathbf{M}_{\ell\ell'}$ computed from the weighted mask. After inverting this relation and binning into linear bins of width $\Delta\ell$, we obtain an unbiased estimate $\hat{C}_b$. The shot‑noise contribution $1/\bar{n}_{\rm eff}$ is subtracted, where $\bar{n}_{\rm eff}$ is the effective number density derived from the weights and the masked sky area.

For two galaxy samples (ETG and LTG) the coupled cross‑spectrum $C_\ell^{ij}$ is formed from the spherical‑harmonic coefficients of the two overdensity maps. The same mode‑coupling formalism is applied using a joint mask, yielding the binned cross‑spectrum $\hat{C}_b^{ij}$. The cross‑spectrum is free of shot‑noise because the two samples consist of disjoint objects. 

As a parametric framework, we can exploit the fact that DESI provides compatible redshift distributions and interpret the ratio between $C_\ell^{ii}$ and $C_\ell^{ij}$ as tracing the ratio of their galaxy biases, which we will name from now on the cross‑tracer clustering ratio (CTCR)\cite{Ferreira:2026iru}
\begin{equation}\label{eq:ratio}
R_\ell^{ij} = \frac{C_\ell^{ii}}{C_\ell^{ij}}.
\end{equation}
\section{The Acoustic Response Estimator}
\label{sec:acoustic_response_estimator}

The Cross-Tracer Clustering Ratio (CTCR) is built from the angular clustering of two morphology-selected galaxy populations. To motivate the estimator introduced in this work, we begin with the galaxy overdensity field of each tracer and derive the expected form of the ratio in the linear and non-linear bias regimes. Throughout, we adopt the notation as in \citet{Ferreira:2026iru} for the empirical scale-dependent bias.

Let \(i \in \{E, L\}\) denote the two tracers, where \(E\) stands for early-type galaxies (ETGs) and \(L\) for late-type galaxies (LTGs). The
three-dimensional galaxy overdensity field of tracer \(i\) can be expanded in a renormalized Eulerian bias expansion
\citep{McDonald2006Bias,DesjacquesEtAl2018}:
\begin{align}\nonumber
    \delta_i(\bm{x})
    &=
    b_{1i}\delta(\bm{x})
    +
    \frac{b_{2i}}{2}[\delta^2](\bm{x})
    +
    b_{s^2 i}[s^2](\bm{x})
    +\nonumber
    \\ &\quad
    b_{\nabla^2\delta,i}
    \frac{\nabla^2\delta(\bm{x})}{k_*^2}
    +
    \sum_{O_3} b_{O_3 i}[O_3](\bm{x})
    +
    \epsilon_i(\bm{x})
    \\ &\quad
    +
    \epsilon_{\delta,i}(\bm{x})\delta(\bm{x})
    +
    \cdots,
    \label{eq:bias_expansion}
\end{align}
where: \(\delta\) is the matter overdensity and \(\nabla^2\delta\) is its Laplacian.
\([\delta^2]\) and \([s^2]\) are the renormalized quadratic density and tidal operators, with the tidal scalar defined as \(s^2 = s_{ij}s_{ij}\) and the tidal tensor given by
    \begin{equation}
        s_{ij}
        =
        \left(
        \frac{\partial_i\partial_j}{\nabla^2}
        -
        \frac{\delta_{ij}}{3}
        \right)\delta.
    \end{equation}
\([O_3]\) denotes the independent cubic operators. \(k_*\) is the reference wavenumber at which the higher-derivative operator \(\nabla^2\delta/k_*^2\) is normalized, typically chosen to be of order the non-linear scale \(k_{\rm NL}\). \(\epsilon_i\) is a stochastic field representing the departure of tracer \(i\) from a deterministic function of the
    matter field, and \(\epsilon_{\delta,i}\) parametrizes a stochastic contribution proportional to \(\delta\). \(b_{1i}\) and \(b_{2i}\) are the linear and quadratic bias
    parameters of tracer \(i\), \(b_{s^2 i}\) is its tidal bias, and   \(b_{\nabla^2\delta,i}\) and \(b_{O_3 i}\) are the coefficients of the higher-derivative and cubic operators respectively.
    
The square brackets indicate renormalized composite operators, for which cutoff-dependent pieces and terms degenerate with lower-order bias or stochastic power have been absorbed into the corresponding parameters. The coefficients in Eq.~\eqref{eq:bias_expansion} are treated as redshift-dependent parameters within each tomographic bin, not as arbitrary functions of scale; the scale dependence is generated by the operator power spectra, higher-derivative terms, and projection effects.

The projected galaxy overdensity in a redshift bin is obtained by integrating along the line of sight,
\begin{equation}
    \Delta_i(\hat{\bm{n}})
    =
    \int dz\, W_i(z)\,
    \delta_i\!\left[\chi(z)\hat{\bm{n}}, z\right],
    \label{eq:projected_overdensity}
\end{equation}
where \(W_i(z)\) is the normalized radial selection function of the bin and \(\chi(z)\) is the comoving distance.

Expanding the projected field in spherical harmonics,
\begin{equation}
    \Delta_i(\hat{\bm{n}})
    =
    \sum_{\ell m} a_{\ell m}^i Y_{\ell m}(\hat{\bm{n}}),
\end{equation}
the angular auto- and cross-power spectra are defined by
\begin{equation}
    \left\langle
    a_{\ell m}^i
    a_{\ell' m'}^{j*}
    \right\rangle
    =
    \delta_{\ell\ell'}\delta_{mm'}
    C_\ell^{ij}.
\end{equation}
Under the Limber approximation, the spectra can be written as
\begin{align}
    C_\ell^{ij}
    =
    \int dz\,
    \frac{H(z)}{c\,\chi^2(z)}
    W_i(z)W_j(z)
    P_{ij}\!\left(k_\ell, z\right),
    \\
    k_\ell = \frac{\ell+1/2}{\chi(z)},
    \label{eq:limber}
\end{align}
where \(P_{ij}(k,z)\) is the three-dimensional cross-power spectrum between tracers \(i\) and \(j\).

The CTCR is defined as the ratio of the ETG auto-spectrum to the ETG--LTG cross-spectrum within the same redshift bin:
\begin{equation}
    \mathcal{R}_\ell
    \equiv
    \frac{C_\ell^{EE}}
    {C_\ell^{EL}}.
    \label{eq:ctcr_ratio}
\end{equation}

In the deterministic linear-bias limit, the galaxy overdensity of each tracer is proportional to the matter field,
\begin{equation}
    \delta_E = b_{1E}\delta,
    \qquad
    \delta_L = b_{1L}\delta,
    \label{eq:linear_bias}
\end{equation}
so that
\begin{equation}
    P_{EE} = b_{1E}^2 P_{\rm lin},
    \qquad
    P_{EL} = b_{1E}b_{1L} P_{\rm lin},
    \label{eq:linear_spectra}
\end{equation}
where \(P_{\rm lin}(k)\) is the linear matter power spectrum. For matched radial kernels, \(W_E(z) \simeq W_L(z) \equiv W(z)\), the Limber integral gives
\begin{equation}
    C_\ell^{EE} \simeq b_{1E}^2 S_\ell,
    \qquad
    C_\ell^{EL} \simeq b_{1E}b_{1L} S_\ell,
\end{equation}
where
\begin{equation}
    S_\ell
    =
    \int dz\,
    \frac{H(z)}{c\,\chi^2(z)}
    W^2(z)
    P_{\rm lin}\!\left(k_\ell, z\right)
\end{equation}
contains the complete projected matter spectrum, including both the smooth broadband shape and the baryon acoustic oscillation (BAO) wiggles. 
Allowing for a scale-dependent cross-correlation coefficient \(r_{EL}(k)\) between the two tracers,
\begin{equation}
    P_{EL}(k) = b_E(k)b_L(k)r_{EL}(k)P_{\rm lin}(k),
\end{equation}
the ratio becomes
\begin{equation}
    \mathcal{R}(k)
    =
    \frac{b_E(k)}{b_L(k)r_{EL}(k)},
    \label{eq:ctcr_with_r}
\end{equation}
which is generally scale dependent. The cross-correlation coefficient \(r_{EL}\) can be measured from the data through
\begin{equation}
    r_{EL,\ell}
    =
    \frac{C_\ell^{EL}}
    {\sqrt{C_\ell^{EE}C_\ell^{LL}}},
    \label{eq:angular_r}
\end{equation}
up to noise and projection effects. The presence of \(r_{EL}(k)\neq 1\) already introduces a smooth scale dependence in the CTCR.

\subsection{Residual acoustic response beyond linear bias}
\label{subsec:non-linear_acoustic_response}

Beyond linear bias, the non-linear operators in Eq.~\ref{eq:bias_expansion} generate additional mode-coupling spectra whose shapes differ from \(P_{\rm lin}(k)\). Because the linear common component is strongly suppressed in the CTCR, differences between the normalized non-linear bias coefficients of ETGs and LTGs can leave a residual scale dependence. Since the mode-coupling spectra are themselves computed from a linear power spectrum containing acoustic oscillations, part of this residual can be BAO dependent.

To display the mechanism explicitly, we retain only the linear and quadratic density operators,
\begin{equation}
    \delta_i
    =
    b_{1i}\delta
    +
    \frac{b_{2i}}{2}[\delta^2].
    \label{eq:quadratic_bias}
\end{equation}
The ETG auto-spectrum is then
\begin{align}
    P_{EE}(k)
    ={}&
    b_{1E}^2 P_{\delta\delta}(k)
    + b_{1E}b_{2E} P_{\delta,[\delta^2]}(k)
    \nonumber\\
    &+
    \frac{b_{2E}^2}{4} P_{[\delta^2],[\delta^2]}(k)
    + P_{\epsilon_E\epsilon_E}(k)
    + \cdots,
    \label{eq:pee_quadratic}
\end{align}
while the ETG--LTG cross-spectrum is
\begin{align}
    P_{EL}(k)
    ={}&
    b_{1E}b_{1L} P_{\delta\delta}(k)
    \nonumber\\
    &+
    \frac{b_{1E}b_{2L} + b_{1L}b_{2E}}{2}
    P_{\delta,[\delta^2]}(k)
    \nonumber\\
    &+
    \frac{b_{2E}b_{2L}}{4} P_{[\delta^2],[\delta^2]}(k)
    + P_{\epsilon_E\epsilon_L}(k)
    + \cdots.
    \label{eq:pel_quadratic}
\end{align}
Neglecting the stochastic terms for clarity and defining
\begin{equation}
    u_i \equiv \frac{b_{2i}}{2b_{1i}},
    \qquad
    \mathcal{X}(k)
    \equiv
    \frac{P_{\delta,[\delta^2]}(k)}{P_{\delta\delta}(k)},
    \qquad
    \mathcal{Y}(k)
    \equiv
    \frac{P_{[\delta^2],[\delta^2]}(k)}{P_{\delta\delta}(k)},
    \label{eq:xy_definitions}
\end{equation}
the three-dimensional CTCR becomes
\begin{equation}
    \mathcal{R}(k)
    =
    \left(\frac{b_{1E}}{b_{1L}}\right)
    \frac{
    1 + 2u_E \mathcal{X}(k) + u_E^2 \mathcal{Y}(k)
    }{
    1 + [u_E + u_L]\mathcal{X}(k) + u_E u_L \mathcal{Y}(k)
    }.
    \label{eq:exact_quadratic_ratio}
\end{equation}
Expanding to first order in the non-linear corrections gives
\begin{align}\label{eq:expanded_quadratic_ratio}\nonumber
    \mathcal{R}(k)
    \simeq
    \frac{b_{1E}}{b_{1L}}\{
    1
    + 
    (u_E - u_L)\mathcal{X}(k)
    +\\
    u_E(u_E - u_L)\mathcal{Y}(k)
    +
    \cdots\}
\end{align}

Equation~\eqref{eq:expanded_quadratic_ratio} shows that a residual survives whenever
\begin{equation}
    \frac{b_{2E}}{b_{1E}}
    \neq
    \frac{b_{2L}}{b_{1L}},
    \label{eq:differential_quadratic_condition}
\end{equation}
i.e.\ whenever the normalized non-linear responses of the two tracers differ.

To see why these terms can contain acoustic information, decompose the linear matter spectrum into smooth and oscillatory parts,
\begin{equation}
    P_{\rm lin}(k)
    =
    P_{\rm nw}(k) + P_{\rm w}(k),
    \label{eq:wiggle_decomposition}
\end{equation}
where \(P_{\rm nw}\) is a no-wiggle spectrum and \(P_{\rm w}\) contains the BAO oscillations. The quadratic-operator spectra are given by
convolution integrals over pairs of linear modes. Writing the convolution schematically as
\begin{align}
    P_{[\delta^2],[\delta^2]}(k)
    &\supset
    2\int\frac{d^3q}{(2\pi)^3}
    P_{\rm lin}(q) P_{\rm lin}(|\bm{k}-\bm{q}|),
    \label{eq:p22_convolution}
    \\
    P_{\delta,[\delta^2]}(k)
    &\supset
    2\int\frac{d^3q}{(2\pi)^3}
    F_2(\bm{q},\bm{k}-\bm{q})
    P_{\rm lin}(q) P_{\rm lin}(|\bm{k}-\bm{q}|),
    \label{eq:pd2_convolution}
\end{align}
where \(P_{\rm lin}(k)\) is the linear matter power spectrum at the redshift of the bin, so that \(P_{\rm lin}(q)\) and \(P_{\rm lin}(|\bm{k}-\bm{q}|)\) are the power spectra of the two     modes \(\bm{q}\) and \(\bm{k}-\bm{q}\) that are being coupled to produce a fluctuation at wave-number \(\bm{k}\). The integral runs over all internal momenta \(\bm{q}\), and the factor of 2 accounts for the two permutations of the two modes in the symmetric operator product. 
\(F_2(\bm{q}_1,\bm{q}_2)\) is the symmetrized second-order gravitational kernel that describes how two linear modes \(\bm{q}_1\) and \(\bm{q}_2\) combine to produce a second-order
    density fluctuation. It is given explicitly by
    \begin{equation}
        F_2(\bm{q}_1,\bm{q}_2)
        =
        \frac{5}{7}
        +
        \frac{1}{2}
        \frac{\bm{q}_1\cdot\bm{q}_2}{q_1 q_2}
        \left(\frac{q_1}{q_2} + \frac{q_2}{q_1}\right)
        +
        \frac{2}{7}
        \frac{(\bm{q}_1\cdot\bm{q}_2)^2}{q_1^2 q_2^2},
        \label{eq:F2_kernel}
    \end{equation}
where \(q_i \equiv |\bm{q}_i|\). The three terms correspond to the three independent ways in which the linear velocity and density fields can combine at second order in perturbation theory.

Substituting Eq.~\eqref{eq:wiggle_decomposition} into Eqs.~\eqref{eq:p22_convolution} and~\eqref{eq:pd2_convolution}
produces terms of four types, proportional to \(P_{\rm nw}P_{\rm nw}\), \(P_{\rm nw}P_{\rm w}\), \(P_{\rm w}P_{\rm nw}\), and
\(P_{\rm w}P_{\rm w}\). The first is the no-wiggle contribution, which would be present even if the BAO feature were absent. The last three
involve at least one factor of the oscillatory part \(P_{\rm w}\). They vanish identically in a matched no-wiggle calculation, in which
the linear spectrum is replaced by \(P_{\rm nw}\) everywhere, and therefore constitute a BAO-dependent non-linear response. Because the convolution mixes modes of different wavelengths, the acoustic oscillations imprinted in \(P_{\rm lin}\) are no longer guaranteed to appear at the same wave-numbers in the quadratic-operator spectra:
non-linear mode coupling can damp the oscillations, shift their phase, or generate out-of-phase acoustic structure
\citep{CrocceScoccimarro2008,IvanovSibiryakov2018}.

A precise definition of the acoustic contribution to the CTCR is therefore
\begin{equation}
    \Delta \mathcal{R}_{\rm BAO}(k)
    \equiv
    \mathcal{R}\!\left[P_{\rm nw}+P_{\rm w}\right](k)
    -
    \mathcal{R}\!\left[P_{\rm nw}\right](k),
    \label{eq:delta_r_bao}
\end{equation}
where the two predictions use the same cosmological broadband parameters, bias coefficients, selection functions, survey window, and nuisance prescription. Using Eq.~\eqref{eq:expanded_quadratic_ratio}, the leading schematic dependence is
\begin{align}\nonumber
    \Delta \mathcal{R}_{\rm BAO}(k)
    \simeq
    \frac{b_{1E}}{b_{1L}}
    (u_E - u_L)
    [
    \Delta \mathcal{X}_{\rm BAO}(k) \\
    + u_E \Delta \mathcal{Y}_{\rm BAO}(k)
    + \cdots
    ],
    \label{eq:delta_r_bao_quadratic}
\end{align}
where
\begin{equation}
    \Delta \mathcal{X}_{\rm BAO}
    =
    \mathcal{X}_{\rm w} - \mathcal{X}_{\rm nw},
    \qquad
    \Delta \mathcal{Y}_{\rm BAO}
    =
    \mathcal{Y}_{\rm w} - \mathcal{Y}_{\rm nw}.
\end{equation}
Equation~\eqref{eq:delta_r_bao_quadratic} expresses the acoustic response as a product of two ingredients: the difference between the
normalized non-linear bias coefficients of the two tracers, \(u_E - u_L\), and the BAO-dependent part of the quadratic operator spectra, \(\Delta \mathcal{X}_{\rm BAO}\) and \(\Delta \mathcal{Y}_{\rm BAO}\). In principle, this is a complete prediction. In practice, it is not directly usable. The non-linear bias coefficients \(b_{2i}\), \(b_{s^2 i}\), and \(b_{\nabla^2\delta,i}\) of ETGs and LTGs are not known a priori, and they depend on halo mass, halo occupation, and assembly history in ways that are not captured by a single number per tracer \cite{Gao_2005, Gao_2007, Dalal2008}. The mapping from the three-dimensional spectra \(P_{ij}(k)\) to the projected angular spectra \(C_\ell^{ij}\) further convolves the acoustic response with the radial selection functions of the two tracers, and redshift-space distortions and the survey window modify both the amplitude and the shape of the residual. Finally, the two tracers are not perfectly correlated, so the CTCR is an effective ratio rather than the pure bias ratio, as we showed in Section~\ref{subsec:r_ell_results}.

For all these reasons, we do not attempt to predict the acoustic response from first principles. Instead, we adopt a phenomenological parametrization that captures the two physical ingredients of Eq.~\eqref{eq:delta_r_bao_quadratic}: a smooth, scale-dependent baseline describing the effective relative bias of the two tracers, and an acoustic modulation whose amplitude \(A_{\rm morph}\) measures the size of the differential non-linear response. The baseline is flexible enough to absorb the unknown non-linear bias coefficients and the projection effects, while the acoustic template is fixed by the BAO physics. This is the same strategy adopted in \cite{Ferreira:2026iru}, where the smooth relative bias of ETGs and LTGs was modelled with a sigmoid function of multipole, and it is the natural extension of that work to the acoustic sector.

\subsection{Empirical scale-dependent bias and the acoustic response estimator}
\label{subsec:estimator_definition}

Following the empirical parametrization of \cite{Ferreira:2026iru}, we model the effective scale-dependent bias of tracer \(i\) as
\begin{equation}
    b_\ell^i
    =
    b_1^i
    +
    \left(b_2^i - b_1^i\right)
    \left[
    1
    +
    \exp\left(
    -\frac{\ell - \ell_0^i}{\Delta^i}
    \right)
    \right]^{-1},
    \label{eq:sigmoid_bias}
\end{equation}
where \(b_1^i\) and \(b_2^i\) are the low- and high-\(\ell\) asymptotes of the sigmoid, \(\ell_0^i\) is the transition scale, and \(\Delta^i\) controls the width of the transition. These parameters are not the perturbative bias coefficients of Eq.~\ref{eq:bias_expansion}; they are effective quantities that absorb the combined scale dependence from non-linear bias, redshift-space distortions, and projection.

With this parametrization, the smooth CTCR baseline is
\begin{equation}
    \mathcal{R}_\ell^{\rm smooth}
    =
    \frac{b_\ell^E}{b_\ell^L}.
    \label{eq:smooth_ctcr}
\end{equation}
Since \(r_\ell\) enters the CTCR as a smooth multiplicative factor, any acoustic modulation of \(r_\ell\) is negligible on the scales
of the BAO. This is not an empirical statement about flatness, but a consequence of the physical origin of \(r_\ell\). The BAO is a feature of
the linear matter power spectrum \(P_m(k)\), which is common to both tracers: all three spectra $C_\ell^{EE}$, $C_\ell^{LL}$ and
$C_\ell^{EL}$ are projections of the same underlying matter field. In the correlation coefficient
\begin{equation}
  r_\ell \;=\; \frac{C_\ell^{EL}}{\sqrt{C_\ell^{EE}\,C_\ell^{LL}}},
\end{equation}
the BAO wiggles therefore cancel at leading order, leaving only a residual modulation controlled by the relative visibility of the
acoustic feature in the two tracers. This residual modulation originates from differential nonlinear damping, which enters through the ratio of
BAO visibilities
\begin{equation}
  \frac{\text{BAO}_{EL}}{\sqrt{\text{BAO}_{EE}\,\text{BAO}_{LL}}}
  \;=\; \exp\!\left[\frac{k^2}{8}\left(\Sigma_E - \Sigma_L\right)^2\right],
  \label{eq:r_damping}
\end{equation}
where \(\Sigma_E\) and \(\Sigma_L\) are the nonlinear damping scales of the two tracers, and \(\Sigma_{EL}\) has been approximated by their mean.
Equation~\eqref{eq:r_damping} is a slowly varying, monotonic function of \(k\) in the acoustic range \(k\!\sim\!0.05\)--\(0.1\,h\,\mathrm{Mpc}^{-1}\)
and contains no oscillatory structure. The same reasoning applies to shot noise, which suppresses \(r_\ell\) monotonically towards high
\(\ell\), and to nonlinear mode decorrelation, which sets in only above \(k\!\sim\!0.2\,h\,\mathrm{Mpc}^{-1}\), well outside the acoustic range.

Motivated by Eq.~\eqref{eq:delta_r_bao_quadratic}, we model the measured CTCR ratio as the product of this smooth baseline and a scale-dependent acoustic modulation:
\begin{equation}
    \mathcal{R}_\ell^{\rm model}
    =
    \mathcal{R}_\ell^{\rm smooth}
    \left[
    1
    +
    A_{\rm morph}\,
    T_\ell^{\rm morph}(\alpha)
    \right],
    \label{eq:acoustic_response_model}
\end{equation}
where the differential acoustic template is defined as
\begin{equation}
    T_\ell^{\rm morph}(\alpha)
    \equiv
    \frac{
    \mathcal{R}_\ell^{\rm w}(\alpha)
    -
    \mathcal{R}_\ell^{\rm nw}
    }{
    \mathcal{R}_\ell^{\rm nw}
    },
    \label{eq:acoustic_template}
\end{equation}
obtained from perturbation theory or from paired wiggle/no-wiggle simulations after applying the same radial kernels, mask, mode-coupling matrix, and band-power windows as in the data.

The acoustic response amplitude \(A_{\rm morph}\) has a direct null interpretation: $A_{\rm morph}$ = 0 means no detected morphology-dependent acoustic response, whereas \(A_{\rm morph} = 1\) corresponds to the fiducial theoretical or simulation-calibrated response. The dilation parameter
\begin{equation}
    \alpha
    \equiv
    \frac{[D_M(z)/r_d]}
    {[D_M(z)/r_d]_{\rm fid}}
    \label{eq:alpha_definition}
\end{equation}
encodes any shift in the acoustic scale relative to the fiducial cosmology.

The likelihood is defined using the full covariance matrix of the ratio,
\begin{equation}
\begin{split}
    \ln \mathcal{L}(\boldsymbol{\Theta})
    &= -\frac{1}{2}
    \left[
    \mathcal{R}_{\rm obs}
    -
    \mathcal{R}_{\rm model}(\boldsymbol{\Theta})
    \right]^{\!\top}
    \mathbf{C}_{\mathcal{R}}^{-1} \\
    &\quad \times
    \left[
    \mathcal{R}_{\rm obs}
    -
    \mathcal{R}_{\rm model}(\boldsymbol{\Theta})
    \right],
\end{split}
\label{eq:likelihood}
\end{equation}
where \(\boldsymbol{\Theta}\) includes the acoustic parameters \(\{A_{\rm morph},\alpha\}\) and the broadband nuisance parameters \(\{b_1^E,b_2^E,\ell_0^E,\Delta^E,b_1^L,b_2^L,\ell_0^L,\Delta^L\}\). The covariance matrix \(\mathbf{C}_{\mathcal{R}}\) is computed from the bootstrap realisations introduced in \cite{Ferreira:2026iru}. Each galaxy is assigned to a \texttt{HEALPix} pixel at \texttt{NSIDE}=512, and for each realisation we randomly select $80\%$ of the occupied pixels with replacement, retaining all galaxies in the chosen pixels. This preserves the large-scale angular correlations that would be destroyed by a galaxy-by-galaxy resampling. We generate $N_{\rm real}=100$ realisations per redshift bin and compute the auto- and cross-spectra with \texttt{NaMaster}. The resulting empirical covariance is regularised by linear shrinkage towards its diagonal with a fixed intensity $\alpha=0.2$, and the inverse is computed with the Hartlap correction to account for the finite number of realisations. The cross-covariance between the auto- and cross-spectra is supplied by the analytic Gaussian covariance of \cite{Ferreira:2026iru}, and we have verified that the full covariance is positive definite for every redshift bin and that its inclusion does not change the CTCR or the bias parameters within the statistical uncertainties.

\begin{figure*}
\centering
\includegraphics[width=0.95\textwidth]{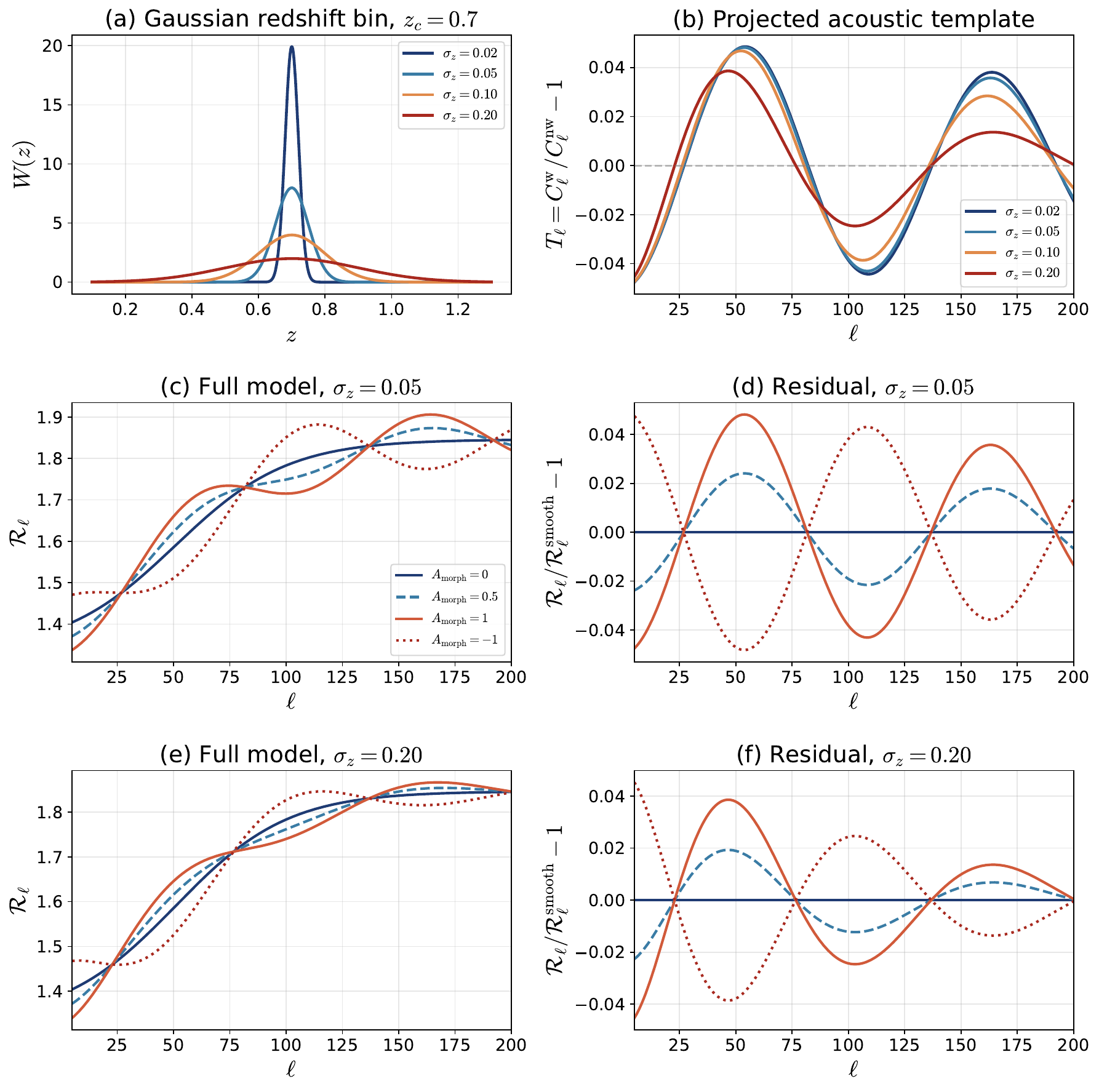}
\caption{
    Acoustic template computed from a Gaussian redshift bin by Limber
    projection of a linear power spectrum with BAO wiggles.
    \textbf{(a)} Gaussian window functions \(W(z)\) centered at
    \(z_c = 0.5\) for four bin widths.
    \textbf{(b)} The projected template
    \(T_\ell = C_\ell^{\rm w}/C_\ell^{\rm nw} - 1\) for the same four
    widths. The number of visible wiggles decreases with increasing
    bin width, because the projection averages over a range of
    transverse wavenumbers at fixed \(\ell\).
    \textbf{(c, d)} Full model and residual for varying \(A_{\rm
    morph}\) at \(\sigma_z = 0.05\).
    \textbf{(e, f)} Same for \(\sigma_z = 0.20\).
}
\label{fig:a_morph_gaussian_bin}
\end{figure*}

Figure~\ref{fig:a_morph_gaussian_bin} shows the acoustic template computed directly from a Gaussian redshift bin by Limber projection of a linear power spectrum containing BAO wiggles. Rather than assuming an analytic form for the template, we take the toy linear spectrum \(P_{\rm lin}(k) = P_{\rm nw}(k)[1 - A_w \cos(k r_d) e^{-(k\sigma_{\rm nl})^2}]\) with \(r_d = 147.1\) Mpc, \(\sigma_{\rm nl} = 8\) Mpc, and \(A_w = 0.05\), project it through the kernel \(W(z) \propto \exp[-(z-z_c)^2/2\sigma_z^2]\) centered at \(z_c = 0.5\) using the Limber approximation, and form the fractional modulation \(T_\ell = C_\ell^{\rm w}/C_\ell^{\rm nw} - 1\). Panel~(a) shows the four Gaussian windows used, with \(\sigma_z = 0.02, 0.05, 0.10,\) and \(0.20\). Panel~(b) shows the resulting templates. The key point is that the number of visible wiggles is controlled by the width of the radial kernel: for the narrowest bin (\(\sigma_z = 0.02\), corresponding to a comoving smearing of roughly \(67\) Mpc at \(z = 0.5\)) three to four oscillations survive, with the first peak at \(\ell \simeq 29\), the second at \(\ell \simeq 58\), and the third near \(\ell \simeq 87\), each progressively damped by the non-linear suppression factor in the power spectrum; for the intermediate bin (\(\sigma_z = 0.05\), smearing of \(\sim 167\) Mpc) only the first two wiggles remain clearly visible; and for the widest bin (\(\sigma_z = 0.20\), smearing of \(\sim 667\) Mpc) only the first peak survives above the noise. This is not because the radial kernel erases the wiggles, but because the projection averages over a range of transverse wavenumbers at fixed \(\ell\), so the oscillation phase varies across the bin and the wiggles are progressively smoothed out. Panels~(c) and~(d) show the full CTCR model and the oscillatory residual at \(\sigma_z = 0.05\) for four values of the acoustic response amplitude \(A_{\rm morph}\), and panels~(e) and~(f) show the same at \(\sigma_z = 0.20\). In both cases the amplitude \(A_{\rm morph}\) simply scales the projected template, and the null hypothesis \(A_{\rm morph} = 0\) corresponds to a flat line in the residual panels. 

The same projection machinery applies directly to the CTCR, which is the central observable of this work. In the matched-kernel deterministic linear-bias limit, both spectra share the same projected matter power spectrum \(S_\ell\), so the ratio reduces to \(\mathcal{R}_\ell^{\rm smooth} = b_\ell^{E}/b_\ell^{L}\), and the BAO wiggles cancel exactly together with the smooth broadband. This is the smooth baseline shown in the top-right panel of Figure~\ref{fig:a_morph_gaussian_bin}, and it is what the sigmoid of Eq.~\eqref{eq:sigmoid_bias} is designed to fit. The residual shown in the bottom-left and bottom-right panels of the same figure is what survives when the cancellation is not perfect: it is proportional to the difference in the normalized nonlinear responses of the two tracers, \(u_E - u_L\), where \(u_i = b_{2i}/(2b_{1i})\), and it is scaled by the amplitude \(A_{\rm morph}\). In this sense, the projected template shown in panel~(b) is not an assumption about the CTCR itself but a prediction for the shape of the residual under a given bin width, and the number of visible wiggles is a direct diagnostic of how well the two tracers are matched radially. When the two kernels \(W_E(z)\) and \(W_L(z)\) are identical, the cancellation is exact, the smooth baseline is recovered, and \(A_{\rm morph} = 0\). When the kernels differ, the residual acquires an oscillatory component whose shape is set by the same projection kernel that governs the number of surviving wiggles in panel~(b), so a narrow bin with well-matched kernels produces a sharper, more localized template, while a wider or mismatched bin produces a broader and more heavily damped one. The figure therefore illustrates both the strength and the limitation of the CTCR as a probe of the acoustic scale: it isolates the differential acoustic response of the two tracer populations, but it does so at the cost of relying on well-matched radial kernels, and the same physics that controls the shape of the projected template also controls the sensitivity of \(A_{\rm morph}\) to the quality of the tracer matching.

\begin{figure*}
\centering
\includegraphics[width=0.95\textwidth]{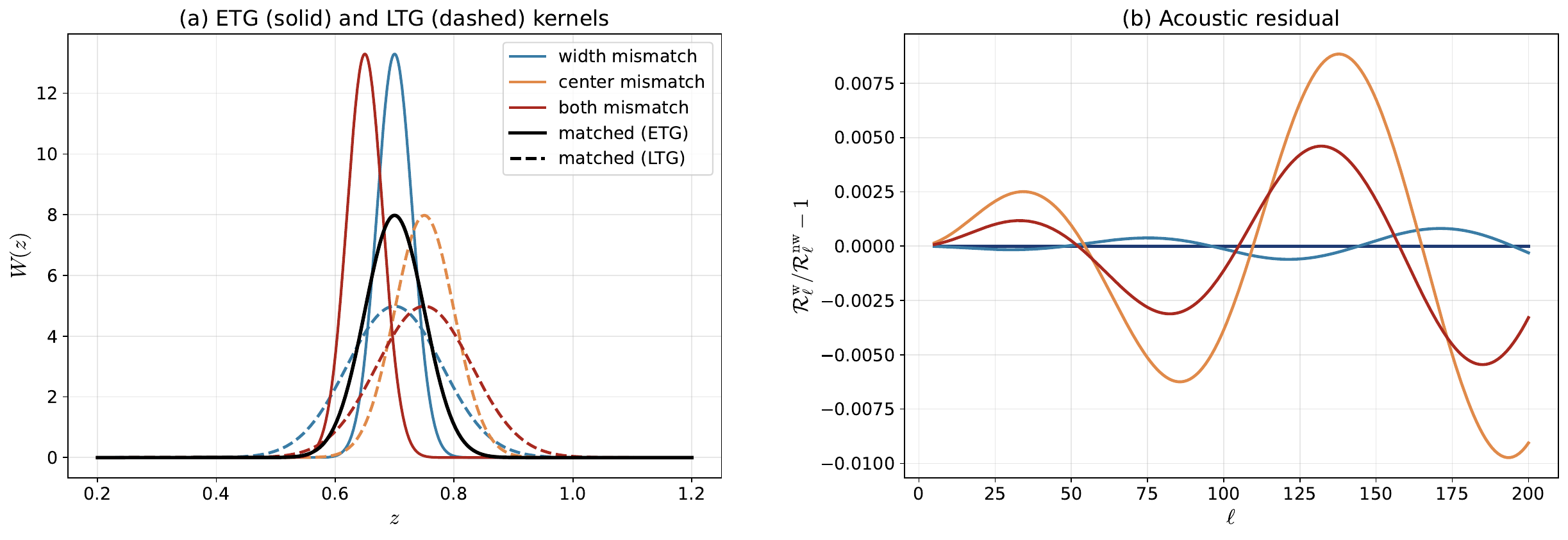}
\caption{
    Effect of ETG--LTG radial kernel mismatch on the CTCR.
    \textbf{(a)} Gaussian redshift kernels for four scenarios,
    all centred at \(z_c = 0.7\). The matched configuration
    (black, drawn on top) uses identical kernels for both tracers
    with \(\sigma_z = 0.05\). The three mismatch scenarios are a
    width mismatch (ETG \(\sigma_z = 0.03\), LTG \(\sigma_z = 0.08\),
    blue), a centre mismatch (LTG displaced by \(\Delta z = +0.05\),
    orange), and a combined mismatch (ETG at \(\Delta z = -0.05\)
    with \(\sigma_z = 0.03\), LTG at \(\Delta z = +0.05\) with
    \(\sigma_z = 0.08\), red). Solid lines are ETG, dashed lines
    are LTG.
    \textbf{(b)} Acoustic residual
    \(\mathcal{R}_\ell^{\rm w}/\mathcal{R}_\ell^{\rm nw} - 1\)
    computed by Limber projection of a linear power spectrum with
    BAO wiggles. The matched case isolates the pure BAO signature:
    a train of damped oscillations with the first peak at
    \(\ell_{\rm BAO} = \pi D_M(z)/r_d\). The mismatched cases show
    the same acoustic structure but with an additional smooth
    broadband offset that grows toward low \(\ell\) and that is
    largest for the combined scenario. This smooth residual is the
    systematic that the sigmoid baseline of
    Eq.~\eqref{eq:sigmoid_bias} is designed to absorb, and it
    illustrates why bins with strongly mismatched kernels cannot
    be used for BAO extraction.
}
\label{fig:kernel_mismatch}
\end{figure*}

Figure~\ref{fig:kernel_mismatch} shows how differences between the ETG and LTG radial kernels propagate into the CTCR. The reference configuration, drawn in black in panel~(a), uses identical Gaussian kernels for both tracers, centred at \(z_c = 0.7\) with \(\sigma_z = 0.05\). Three mismatch scenarios are shown for comparison: a width mismatch, in which the ETG kernel is narrower (\(\sigma_z = 0.03\)) and the LTG kernel broader (\(\sigma_z = 0.08\)) at the same centre; a centre mismatch, in which the LTG kernel is displaced by \(\Delta z = +0.05\) relative to the ETG kernel; and a combined scenario, in which the ETG kernel is displaced downward by \(\Delta z = -0.05\) with \(\sigma_z = 0.03\) while the LTG kernel is displaced upward by \(\Delta z = +0.05\) with \(\sigma_z = 0.08\). Panel~(b) shows the resulting acoustic residual \(\mathcal{R}_\ell^{\rm w}/\mathcal{R}_\ell^{\rm nw} - 1\), computed by Limber projection of a linear power spectrum with BAO wiggles onto each kernel pair, where \(\mathcal{R}_\ell^{\rm w}\) and \(\mathcal{R}_\ell^{\rm nw}\) are the CTCR ratios computed with and without the BAO feature respectively. The matched case isolates the pure BAO signature: a train of damped oscillations whose first peak sits at \(\ell_{\rm BAO} = \pi D_M(z)/r_d\) and whose higher wiggles are progressively suppressed by the projection over the finite kernel width and by the nonlinear damping factor. The mismatched cases show the same acoustic structure but with an additional broadband offset. The width mismatch produces a smooth curvature that grows toward low \(\ell\), the centre mismatch produces a monotonic tilt, and the combined scenario produces the largest deviation because it introduces both effects simultaneously. This residual is the systematic that the smooth sigmoid baseline of Eq.~\eqref{eq:sigmoid_bias} is designed to absorb, and it is why the CTCR acoustic response estimator remains unbiased for well-matched tracers but cannot extract a reliable BAO scale when the two kernels differ too strongly: once the smooth residual from the mismatch becomes comparable to the BAO wiggle amplitude, the acoustic template and the broadband baseline become degenerate, and the fit can no longer separate them.

\section{Results}
\label{sec:results}

We now present the CTCR measurements and the constraints on the acoustic response estimator from the DESI Legacy Survey DR8 data and a Euclid Flagship 2 validation. The analysis follows the pipeline described in Section~\ref{sec:acoustic_response_estimator}.

\subsection{BAO dilation parameter}
\label{subsec:alpha_results}

Table~\ref{tab:desi_alpha} summarises the per-bin measurements of the dilation parameter \(\alpha\) and the acoustic response amplitude \(A_{\rm morph}\). The combined constraint on the dilation parameter, obtained by inverse-variance weighting over the nine bins, is
\begin{equation}
    \alpha^{\rm DESI}
    =
    0.9997 \pm 0.0383.
    \label{eq:alpha_desi_combined}
\end{equation}
This is consistent with the Planck 2018 \(\Lambda\)CDM prediction (\(\alpha = 1\)) at the \(0.01\sigma\) level. The per-bin values are all consistent with unity, with individual uncertainties of approximately \(0.115\), reflecting the photometric precision of the DESI Legacy Survey sample.

\subsection{Acoustic response amplitude}
\label{subsec:am_desi_results}

The combined acoustic response amplitude, obtained by inverse-variance weighting over the nine bins, is
\begin{equation}
    A_{\rm morph}^{\rm DESI}
    =
    0.369 \pm 0.307.
    \label{eq:am_desi_combined}
\end{equation}
This is consistent with the null hypothesis \(A_{\rm morph} = 0\) at the \(1.2\sigma\) level. There is no statistically significant evidence for a morphology-dependent acoustic response in the DESI data at the current level of precision.

\begin{table}[htbp]
\centering
\caption{
    CTCR acoustic response measurements from the DESI Legacy Survey DR8 data.
    For each redshift bin we report the effective redshift \(z_{\rm eff}\),
    the acoustic response amplitude \(A_{\rm morph}\), and the dilation parameter
    \(\alpha\). Errors are the standard deviations of the marginalised posterior
    distributions, corresponding to \(68\%\) credible intervals.
    The last row gives the inverse-variance-weighted combined values.
}
\label{tab:desi_alpha}
\begin{tabular}{c c c c}
\hline\hline
Bin & \(z_{\rm eff}\) & \(A_{\rm morph}\) & \(\alpha\) \\
\hline
0 & 0.1189 & \(0.603 \pm 0.997\) & \(1.0004 \pm 0.1145\) \\
1 & 0.2180 & \(0.567 \pm 0.988\) & \(0.9979 \pm 0.1149\) \\
2 & 0.3135 & \(0.275 \pm 0.868\) & \(1.0021 \pm 0.1154\) \\
3 & 0.4127 & \(0.339 \pm 0.916\) & \(0.9979 \pm 0.1142\) \\
4 & 0.5092 & \(0.213 \pm 0.841\) & \(0.9967 \pm 0.1161\) \\
5 & 0.6104 & \(0.313 \pm 0.932\) & \(1.0019 \pm 0.1157\) \\
6 & 0.7063 & \(0.358 \pm 0.907\) & \(1.0009 \pm 0.1144\) \\
7 & 0.8099 & \(0.359 \pm 0.931\) & \(0.9976 \pm 0.1143\) \\
8 & 0.9063 & \(0.401 \pm 0.942\) & \(1.0019 \pm 0.1151\) \\
\hline
\multicolumn{2}{c}{Combined} & \(0.369 \pm 0.307\) & \(0.9997 \pm 0.0383\) \\
\hline\hline
\end{tabular}
\end{table}

The per-bin values, listed in Table~\ref{tab:desi_alpha}, range from \(A_{\rm morph} = 0.21 \pm 0.84\) (at \(z = 0.509\)) to \(A_{\rm morph} = 0.60 \pm 1.00\) (at \(z = 0.119\)), with individual uncertainties of order unity. All nine bins are consistent with zero within \(1\sigma\). The scatter of the per-bin values is consistent with the estimated uncertainties, and there is no systematic trend with redshift.

The acoustic response estimator of Eq.~\eqref{eq:acoustic_response_model} adds two parameters to the smooth sigmoid baseline: the acoustic amplitude \(A_{\rm morph}\) and the dilation parameter \(\alpha\). To determine whether the data require this additional complexity, we compare the smooth-only model
(8 parameters) with the full model (10 parameters) using Bayesian model selection criteria.

Table~\ref{tab:model_comparison_pymc} reports the information criteria and cross-validation between the two models. 
To compare the smooth-only and full models we use four information criteria. All four balance goodness of fit against model complexity,
and lower values indicate a preferred model. The \emph{Akaike Information Criterion} (AIC) and the \emph{Bayesian Information Criterion} (BIC) are defined as
\begin{align}
    {\rm AIC} &= -2\ln\mathcal{L}(\hat{\boldsymbol{\Theta}}) + 2p, \\
    {\rm BIC} &= -2\ln\mathcal{L}(\hat{\boldsymbol{\Theta}}) + p\ln n,
\end{align}
where \(\ln\mathcal{L}(\hat{\boldsymbol{\Theta}})\) is the maximum log-likelihood over the posterior samples, \(p\) is the number of free
parameters, and \(n\) is the number of data points. Both penalise complexity, but BIC penalises it more strongly because the penalty
grows with \(\ln n\).

The \emph{Watanabe--Akaike Information Criterion} (WAIC) is fully
Bayesian and is computed from the posterior predictive distribution \cite{Watanabe2021}. For a data vector of length \(n\)
and a posterior sample \(\{\boldsymbol{\Theta}^{(s)}\}_{s=1}^{S}\), the
WAIC is defined as
\begin{equation}
    {\rm WAIC}
    =
    -2\,\widehat{\rm elpd}_{\rm waic},
\end{equation}
where
\begin{align}
    \widehat{\rm elpd}_{\rm waic}
    =
    \sum_{d=1}^{n}
    \ln\left(
    \frac{1}{S}\sum_{s=1}^{S}
    \mathcal{L}_d(\boldsymbol{\Theta}^{(s)})
    \right)\\ \nonumber
    -
    \sum_{d=1}^{n}
    \widehat{\rm Var}_s\left[
    \ln \mathcal{L}_d(\boldsymbol{\Theta}^{(s)})
    \right],
\end{align}
with \(\mathcal{L}_d(\boldsymbol{\Theta}^{(s)})\) the likelihood of data point \(d\) evaluated at the \(s\)-th posterior draw. The first term measures the goodness of fit, and the second term is the effective number of parameters, which penalises models whose predictions are sensitive to individual posterior draws. 

For the comparison between the smooth-only model (\(p = 8\)) and the full model (\(p = 10\)) we report the differences
\begin{equation}
    \Delta{\rm WAIC}
    =
    {\rm WAIC}_{\rm smooth} - {\rm WAIC}_{\rm full},
\end{equation}
and analogously for \(\Delta{\rm BIC}\) and \(\Delta{\rm AIC}\).
Negative values favour the full model, since the full model has lower WAIC, BIC, and AIC. The results are collected in Table~\ref{tab:model_comparison_pymc}. The table has six columns: the effective redshift \(z_{\rm eff}\), the reduced chi-squared of the smooth-only model and of the full model, and the three information
criteria differences.

The reduced chi-squared values, evaluated at the posterior median parameters, are close to unity for both models in every redshift bin,
ranging from \(0.32\) to \(1.04\) for the smooth-only model and from \(0.43\) to \(1.03\) for the full model. This confirms that both
models provide an adequate description of the measured ratio at the level of the bootstrap uncertainties, and that the extra flexibility
of the acoustic template does not degrade the quality of the fit. The combined values across the nine bins, listed in the last row of
the table, are \(\chi^2/\nu_{\rm smooth} = 0.65\) and \(\chi^2/\nu_{\rm full} = 0.73\).

The three information criteria give consistent results. The full model is marginally preferred in six of the nine redshift bins, with
\(\Delta{\rm WAIC}\) values ranging from \(-0.6\) to \(-1.6\) and significances of at most \(2.6\sigma\). In the remaining three bins
(\(z = 0.31\), \(0.51\), and \(0.81\)), the two models are indistinguishable, with \(\Delta{\rm WAIC} = 0.00 \pm 0.00\).

\begin{table}[htbp]
\centering
\setlength{\tabcolsep}{4pt}
\caption{Bayesian model comparison in the ratio space. The
smooth-only model has \(p = 8\) parameters; the full model adds the
acoustic response amplitude $A_{\rm morph}$ and dilation
$\alpha$, for a total of \(p = 10\). Negative $\Delta$WAIC,
$\Delta$BIC, and $\Delta$AIC favour the full model. The reduced
chi-squared is evaluated at the posterior median parameters. The
last row gives the combined values across the nine bins.}
\label{tab:model_comparison_pymc}
\begin{tabular}{@{}c c c c c c@{}}
\hline\hline
$z_{\rm eff}$ & $\chi^2_{\rm smooth}/\nu$ & $\chi^2_{\rm full}/\nu$
              & $\Delta$WAIC & $\Delta$BIC & $\Delta$AIC \\
\hline
0.1189 & 0.47 & 0.57 & $-0.62 \pm 0.29$ & $-5.21$ & $-3.43$ \\
0.2180 & 0.32 & 0.43 & $-1.46 \pm 0.56$ & $-5.85$ & $-4.07$ \\
0.3135 & 1.01 & 0.83 & $\phantom{-}0.00 \pm 0.00$ & $-1.57$ & $\phantom{-}0.21$ \\
0.4127 & 0.69 & 0.92 & $-1.33 \pm 0.50$ & $-5.65$ & $-3.86$ \\
0.5092 & 1.04 & 1.03 & $\phantom{-}0.00 \pm 0.00$ & $-3.04$ & $-1.26$ \\
0.6104 & 0.79 & 0.99 & $-1.62 \pm 0.55$ & $-4.94$ & $-3.16$ \\
0.7063 & 0.53 & 0.67 & $-1.27 \pm 0.75$ & $-5.96$ & $-4.18$ \\
0.8099 & 0.60 & 0.53 & $\phantom{-}0.00 \pm 0.00$ & $-4.44$ & $-2.66$ \\
0.9063 & 0.44 & 0.57 & $-1.31 \pm 0.53$ & $-5.77$ & $-3.99$ \\
\hline
\multicolumn{6}{@{}l@{}}{\footnotesize Combined: $\chi^2/\nu_{\rm smooth} = 0.65$, $\chi^2/\nu_{\rm full} = 0.73$} \\
\hline\hline
\end{tabular}
\end{table}

The \(\Delta\)BIC values in Table~\ref{tab:model_comparison_pymc} are systematically more negative than the \(\Delta\)WAIC values because
the BIC penalty for two extra parameters, \(p\ln n \approx 6\)--\(7\) for \(n \sim 20\)--\(36\), is comparable to the \(\chi^2\) improvement. With such a small number of data points per bin, BIC is not a reliable model selection criterion, and we do not use it for any quantitative statement. The WAIC criterion, which is based on the full posterior predictive distribution, is more appropriate here.

The acoustic response amplitude \(A_{\rm morph}\) is not the only quantity constrained by the CTCR. The smooth broadband baseline \(\mathcal{R}_\ell^{\rm smooth}\), parametrised by the sigmoid of Eq.~\eqref{eq:sigmoid_bias}, carries information about the scale-dependent relative bias between ETGs and LTGs. The DESI data are consistent with a mild scale dependence, with the transition scale and width constrained to the fitted multipole range in the intermediate-redshift bins. The low- and high-\(\ell\) asymptotes of
the sigmoid are consistent with the values reported in \cite{Ferreira:2026iru}, confirming that the CTCR measurement is stable across the two analyses.

The combination of the two constraints---a BAO scale consistent with Planck 18 \cite{2020} and a null acoustic response---is the central result of this
work. The CTCR recovers the acoustic scale without bias, and the residual morphology-dependent acoustic response is consistent with
zero. This is the expected behaviour for two morphology-selected samples with well-matched redshift distributions, and it demonstrates
that the CTCR method is a robust tool for BAO extraction from photometric data.

\subsection{Cross-correlation coefficient and the effective bias ratio}
\label{subsec:r_ell_results}

The CTCR ratio \(\mathcal{R}_\ell^{EL} = C_\ell^{EE}/C_\ell^{EL}\) is not, in general, equal to the ratio of the two biases. It is related to the true bias ratio through the cross-correlation coefficient
\begin{equation}
    r_\ell
    \equiv
    \frac{C_\ell^{EL}}
    {\sqrt{C_\ell^{EE} C_\ell^{LL}}},
    \label{eq:r_ell_desi}
\end{equation}
which measures the degree to which the two morphological populations trace the same projected large-scale structure. Under the idealised assumptions of deterministic linear bias and perfect correlation, \(r_\ell = 1\) and \(\mathcal{R}_\ell^{EL} = b_E/b_L\). The CTCR measured from the data is therefore an effective ratio that is bounded above by the true bias ratio, with equality only when \(r_\ell = 1\).

\begin{figure}
\centering
\includegraphics[width=0.5\textwidth]{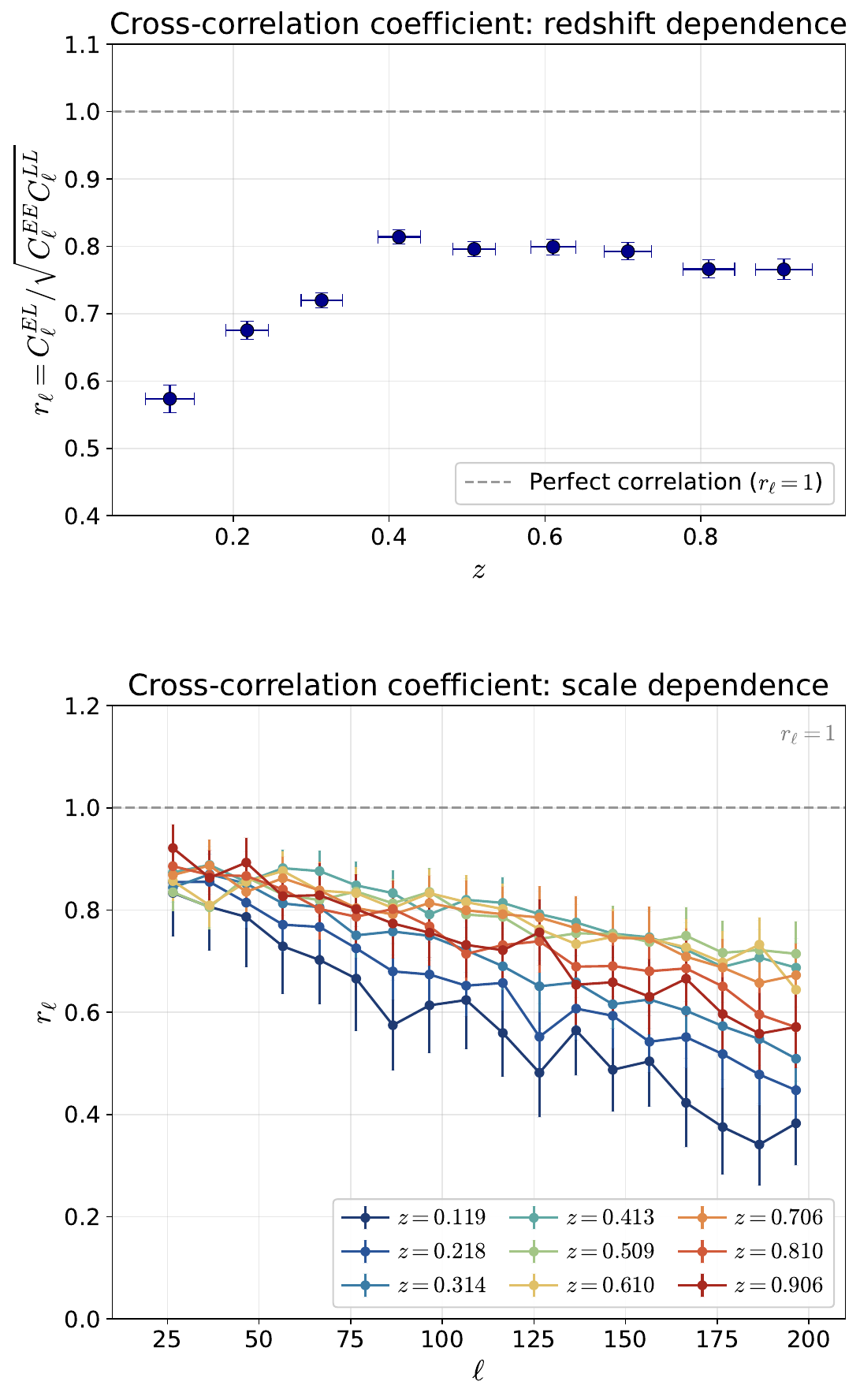}
\caption{
    Cross-correlation coefficient \(r_\ell = C_\ell^{EL} /
    \sqrt{C_\ell^{EE} C_\ell^{LL}}\) from the DESI Legacy Survey DR8
    data.
    \textbf{Top panel:} \(r_\ell\) averaged over the fitted multipole
    range \(20 < \ell < 200\), as a function of redshift. The dashed
    line marks perfect correlation, \(r_\ell = 1\). The coefficient is
    significantly below unity in every bin, with deviations ranging
    from \(15.5\sigma\) to \(25.8\sigma\).
    \textbf{Bottom panel:} \(r_\ell\) as a function of multipole for
    each redshift bin, colour-coded from low (dark) to high (light)
    redshift. The coefficient is consistent with a constant in the
    intermediate-redshift bins (\(0.4 < z < 0.8\)), where the reduced
    chi-squared relative to a constant is \(0.9\)--\(2.2\). In the
    lowest- and highest-redshift bins, \(r_\ell\) shows additional
    scale dependence, with \(\chi^2/\nu\) up to \(5.9\), consistent
    with higher shot noise and less well-matched radial kernels in
    those bins.
}
\label{fig:r_ell_desi}
\end{figure}

Figure~\ref{fig:r_ell_desi} shows the measured cross-correlation coefficient as a function of redshift (top panel) and multipole (bottom panel) for the DESI Legacy Survey data. Table~\ref{tab:r_ell_desi} lists the values averaged over \(20 < \ell < 200\). The coefficient is significantly less than unity in every redshift bin, with per-bin deviations ranging from \(15.5\sigma\) to \(25.8\sigma\), and a combined inverse-variance constraint of \(r_\ell = 0.761 \pm 0.004\). The lowest-redshift bin has the smallest value, \(r = 0.574 \pm 0.021\), while the intermediate-redshift bins cluster around \(r \simeq 0.79\)--\(0.81\). There is a clear redshift trend: the two morphological populations
become progressively more correlated with increasing redshift, and by \(z \sim 0.8\) the correlation stabilises around \(r \simeq 0.77\).

The bottom panel of Fig.~\ref{fig:r_ell_desi} shows that the scale dependence of \(r_\ell\) is not uniform across redshift. We quantify this by computing the reduced chi-squared of \(r_\ell\) relative to a constant in each bin, which is listed in the last column of Table~\ref{tab:r_ell_desi}. In the intermediate-redshift bins (\(0.4 < z < 0.8\)), the values are \(\chi^2/\nu = 0.89\)--\(2.17\), consistent with a scale-independent coefficient. In the lowest-redshift bins (\(z < 0.4\)) and the highest-redshift bin
(\(z \sim 0.9\)), the values are \(\chi^2/\nu = 2.7\)--\(5.9\), indicating significant scale dependence. The scatter of \(r_\ell\) across the fitted \(\ell\) range is \(0.05\)--\(0.07\) in the intermediate bins and \(0.11\)--\(0.15\) in the low- and high-redshift bins. The scale dependence at the extremes of the redshift range is
consistent with the combined effect of higher shot noise and less well-matched radial kernels, both of which are expected to be largest in those bins.

The fact that \(r_\ell < 1\) is physically meaningful. Ellipticals and spirals are not drawn from the same halo population. Ellipticals preferentially occupy more massive, earlier-forming halos, while spirals are more commonly found in lower-mass, later-forming halos. At fixed redshift, the two populations therefore sample different modes of the underlying density field, and their cross-correlation is suppressed relative to the auto-correlations. This is a direct consequence of the different halo occupation statistics of the two morphological types, consistent with the halo assembly bias framework
discussed in \cite{Ferreira:2026iru}.

The bias ratio inferred from the CTCR alone is therefore systematically higher than the true bias ratio. In the intermediate-redshift bins, where \(r_\ell\) is well constrained and approximately scale independent, this correction is of order \(20\%\). The true bias ratio is therefore within the range \(b_E/b_L \sim 1.2\)--\(1.7\) across the DESI sample, consistent with the linear bias measurements of \cite{Ferreira:2026iru} and with the expectation from halo occupation modelling.

\begin{table}[htbp]
\centering
\caption{
    Cross-correlation coefficient \(r_\ell\) for the DESI Legacy Survey
    DR8 data, averaged over the fitted multipole range
    \(20 < \ell < 200\). The fourth column reports the deviation of
    \(r\) from unity in units of the standard error of the mean. The
    fifth column reports the reduced chi-squared of \(r_\ell\) relative
    to a constant in each bin, which quantifies the scale dependence
    visible in the bottom panel of Fig.~\ref{fig:r_ell_desi}. The
    coefficient is significantly below unity in every bin, confirming
    that the CTCR is an effective ratio rather than the physical bias
    ratio \(b_E/b_L\).
}
\label{tab:r_ell_desi}
\begin{tabular}{c c c c c}
\hline\hline
Bin & \(z_{\rm eff}\) & \(\langle r_\ell \rangle\)
    & Deviation & \(\chi^2/\nu\) \\
\hline
0 & 0.1189 & \(0.5737 \pm 0.0208\) & \(-20.5\sigma\) & \(3.06\) \\
1 & 0.2180 & \(0.6752 \pm 0.0132\) & \(-24.6\sigma\) & \(5.51\) \\
2 & 0.3135 & \(0.7197 \pm 0.0108\) & \(-25.8\sigma\) & \(5.85\) \\
3 & 0.4127 & \(0.8137 \pm 0.0105\) & \(-17.8\sigma\) & \(2.17\) \\
4 & 0.5092 & \(0.7958 \pm 0.0114\) & \(-17.9\sigma\) & \(0.89\) \\
5 & 0.6104 & \(0.7991 \pm 0.0117\) & \(-17.2\sigma\) & \(1.44\) \\
6 & 0.7063 & \(0.7925 \pm 0.0128\) & \(-16.2\sigma\) & \(1.55\) \\
7 & 0.8099 & \(0.7662 \pm 0.0135\) & \(-17.4\sigma\) & \(2.68\) \\
8 & 0.9063 & \(0.7656 \pm 0.0151\) & \(-15.5\sigma\) & \(3.07\) \\
\hline\hline
\end{tabular}
\end{table}

This result also clarifies the interpretation of the acoustic response amplitude \(A_{\rm morph}\). The estimator of Section~\ref{subsec:estimator_definition} fits the CTCR ratio with a smooth baseline \(\mathcal{R}_\ell^{\rm smooth}\) plus an acoustic template. Because \(r_\ell < 1\), the smooth baseline is not simply \(b_E/b_L\) but the effective ratio \(\mathcal{R}_\ell^{EL} = (b_E/b_L)/r_\ell\). The acoustic response amplitude then measures whether the residual scale dependence of the effective ratio has an acoustic component. The fact that the DESI data give \(A_{\rm morph} = 0.37 \pm 0.31\), consistent with zero, indicates that the acoustic modulation of the effective ratio is small, while the smooth part absorbs the \(r_\ell\)-driven suppression. This is the expected behaviour if the imperfect correlation is dominated by a smooth, scale-independent effect, as the perturbative expansion of Eq.~\ref{eq:acoustic_response_model} predicts.

\section{Euclid Flagship 2 validation}
\label{subsec:euclid_validation}

To validate the acoustic response estimator and to test whether the null result reported for the DESI morphology-selected sample is a
physical property of the tracers or an artefact of the pipeline, we applied the full CTCR analysis to the Euclid Flagship 2
(FS2) dark-matter-only light cone \citep{Euclid:2024few}. The FS2 mock has a known input cosmology, no intrinsic morphology-dependent
acoustic response, and no observational systematics. Any acoustic response recovered from the mock must therefore originate from the
selection of the tracer populations, not from a physical differential coupling to the BAO.

We define two families of tracers in the FS2 catalogue. The first is
morphology-based, using the \texttt{dominant\_shape} classification
of each galaxy:
\begin{align}
    \text{Elliptical}: &\quad \texttt{dominant\_shape} = 0, \\
    \text{Spiral}:     &\quad \texttt{dominant\_shape} = 1.
\end{align}
The second is mass-based, using the stellar mass
\(\log_{10}(M_\star/M_\odot)\):
\begin{align}
    \text{LTG}:  &\quad \log_{10}(M_\star/M_\odot) < 9.0, \\
    \text{ETG}:  &\quad 9.0 \le \log_{10}(M_\star/M_\odot) < 11.0, \\
    \text{ETG2}: &\quad 9.5 \le \log_{10}(M_\star/M_\odot) < 10.0, \\
    \text{LRG}:  &\quad \log_{10}(M_\star/M_\odot) \ge 11.0, \\
    \text{LRG2}: &\quad \log_{10}(M_\star/M_\odot) \ge 10.5.
\end{align}
The morphology-based pair (Elliptical vs Spiral) is the closest analog to the DESI analysis: two classes of galaxies with similar
redshift distributions and well-matched radial kernels. The mass-based pairs span a range of bias contrasts, from the
strongly separated LRG2/LTG pair to the more closely matched ETG/LTG pair.

For each tracer we select a peaked redshift bin centred on \(z_{\rm peak}\) with width \(\sigma_z = 0.05\), using the
probabilistic selection of \citet{Ferreira:2026iru}. The wide footprint spans \(120^\circ \le \text{RA} \le 260^\circ\) and
\(-30^\circ \le \text{Dec} \le 30^\circ\), and the redshift window is \(0.30 \le z \le 1.20\). Each peaked-bin catalogue is capped at
\(N = 10^6\) galaxies by random subsampling to equalise the statistical weight of the different samples. We compute the
auto- and cross-power spectra with \texttt{NaMaster} \citep{Alonso_2019} at \(\texttt{NSIDE} = 256\) and \(\Delta \ell = 20\), and estimate
the covariance of the CTCR ratio \(R_\ell = C_\ell^{EE}/C_\ell^{EL}\) with a \(4 \times 4\) jackknife over the footprint, regularised
with linear shrinkage and PSD projection. The acoustic response model of Eq.~\eqref{eq:acoustic_response_model} is then fitted to
each bin with \texttt{PyMC} NUTS, using the same priors as the DESI analysis.

\begin{figure*}[htbp]
\centering
\includegraphics[width=0.95\textwidth]{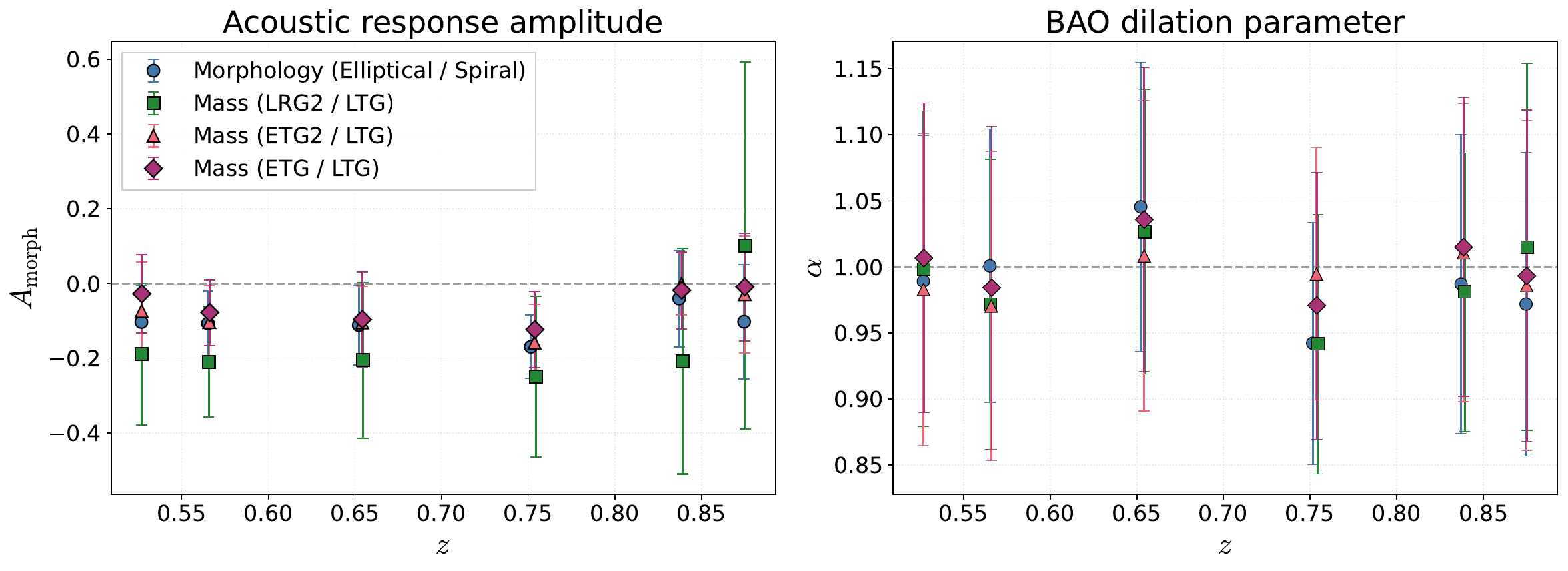}
\caption{
    CTCR acoustic response estimator applied to the Euclid
    Flagship 2 mock, for four tracer pairs and six redshift bins.
    \textbf{Left panel:} acoustic response amplitude
    \(A_{\rm morph}\) as a function of redshift. The dashed line
    marks the null hypothesis \(A_{\rm morph} = 0\). The
    morphology-selected pair (blue circles) is consistent with
    zero across the full redshift range, while the mass-selected
    pairs show a mild negative response at intermediate redshifts
    where the \(n(z)\) mismatch is largest.
    \textbf{Right panel:} dilation parameter \(\alpha\) as a
    function of redshift. The dashed line marks the input
    cosmology (\(\alpha = 1\)). The morphology-selected pair is
    consistent with unity in every bin, while the mass-selected
    pairs show a downward shift of order \(1\sigma\) in the
    intermediate-redshift bins. All error bars are \(68\%\)
    credible intervals from the PyMC NUTS posterior.
}
\label{fig:euclid_ctcr_combined}
\end{figure*}

Figure~\ref{fig:euclid_ctcr_combined} shows the acoustic response amplitude \(A_{\rm morph}\) and the dilation parameter \(\alpha\)
as a function of redshift for the four tracer pairs. The morphology-selected pair (blue circles) gives
consistent with zero and with the input cosmology across all six redshift bins. This is the expected outcome for two tracers with
well-matched radial kernels and no intrinsic differential acoustic response, and it confirms that the DESI null result is a physical
property of the morphology-selected sample, not a limitation of the estimator.

The mass-selected pairs tell a more nuanced story. The ETG2/LTG and ETG/LTG pairs, whose \(n(z)\) shapes differ more strongly from the
LTG reference, show a mild negative acoustic response in the intermediate-redshift bins with a downward shift in \(\alpha\) of order \(1\sigma\). At \(z = 0.75\) the effect grows to \(A_{\rm morph} = -0.17 \pm 0.08\) for the morphology pair and \(-0.25 \pm 0.22\) for LRG2/LTG, though the latter is dominated by the small number of LRG2 objects at this redshift. At \(z = 0.85\) and \(z = 0.95\) the mass-selected pairs return to consistency with zero, while the morphology pair remains null.

The key features of Figure~\ref{fig:euclid_ctcr_combined} can be summarised as follows. The morphology-selected pair is null across the full redshift range: the blue circles in both panels lie on the reference lines \(A_{\rm morph} = 0\) and \(\alpha = 1\) in every bin, with uncertainties of order \(0.10\), making this the direct Euclid FS2 counterpart of the DESI analysis. The mass-selected pairs, by contrast, show a mild negative response in the intermediate-redshift bins, with the effect largest where the \(n(z)\) mismatch between the two tracers is largest and disappearing where the kernels are better matched. The LRG2/LTG pair exhibits the largest scatter, as expected given that the LRG2 sample contains fewer galaxies than the other tracers and therefore has higher shot noise, which limits how well the acoustic response can be constrained. Importantly, no tracer pair shows a statistically significant positive response: all recovered amplitudes are either consistent with zero or mildly negative, consistent with the interpretation that any residual scale dependence in the CTCR arises from kernel mismatch rather than from a physical differential coupling to the BAO.

The Euclid FS2 validation therefore confirms the central result of this work: the acoustic response estimator is sensitive to the quality of the tracer matching, and it returns a null result when the two tracers are drawn from the same underlying population with matched redshift distributions. The morphology-selected pair in Euclid FS2 is the analog of the DESI ETG/LTG sample, and both give \(A_{\rm morph} \simeq 0\) and \(\alpha \simeq 1\).

\subsection{Photo-$z$ robustness of the acoustic response estimator}
\label{sec:photoz_robustness}
The DESI Legacy Survey analysis of Section~\ref{sec:results} relies on photometric redshifts whose uncertainties are $\sigma_z \approx 0.03 - 0.04 (1+z)$, an order of magnitude larger than the spectroscopic precision of the LRG samples used in the DESI DR2 BAO measurement. Two distinct effects could, in principle, contaminate the CTCR acoustic response estimator. First, the radial smearing from photo-$z$ errors broadens the effective kernels $W_{E}(z)$ and $W_{L}(z)$ that enter the Limber projection of Eq.~\eqref{eq:limber}, and a residual scale dependence could survive the smooth sigmoid baseline and be misinterpreted as an acoustic modulation. Second, a systematic calibration bias between the ETG and LTG photo-$z$ distributions would displace one kernel relative to the other by $\Delta\chi \approx \Delta z \times d\chi/dz$, and the resulting mismatch of the two radial windows could mimic a differential response of the two morphological populations. We test both effects directly on the Euclid Flagship~2 mock, which has a known input cosmology and no intrinsic morphology-dependent acoustic response. For each of the four redshift bins already used in Section~\ref{subsec:euclid_validation}, we select the morphology-selected ETG/LTG pair on true redshifts and then apply a photo-$z$ perturbation of the form
\begin{equation}
z \;\longrightarrow\; z + o\,(1+z) + \sigma_{z}\,(1+z)\,\mathcal{N}(0,1) ,
\label{eq:photoz_perturbation}
\end{equation}
where $\sigma_{z}$ is a Gaussian scatter in units of $(1+z)$ and $o$ is a systematic fractional offset. Two families of perturbations are considered. The \emph{scatter} family applies $\sigma_{z}$ to both tracers, either symmetrically ($\sigma_{z}^{E} = \sigma_{z}^{L} = 0.03$) or asymmetrically ($\sigma_{z}^{E}, \sigma_{z}^{L} = 0.02, 0.04$ and $0.01, 0.05$), and isolates the effect of radial smearing alone. The \emph{shift} family applies a systematic offset to one tracer only, on top of a common $\sigma_{z} = 0.03$ scatter, and isolates a calibration bias between the two populations: the LTG kernel is displaced by $+1\%$, $+2\%$, $+3\%$ in $(1+z)$, while the ETG kernel is displaced by the same amounts. The reference configuration uses true redshifts with no perturbation. At the effective redshift of the lowest bin ($z_{\rm eff} \simeq 0.51$), the symmetric $3\%$ scatter corresponds to an rms radial smearing of $\pm 114~\mathrm{Mpc}$, the $2\%$ and $3\%$ systematic shifts to $+100$ and $+150~\mathrm{Mpc}$ respectively, and the $\sigma_{z} = 0.05$ tail of the largest differential-scatter configuration to $\pm 190~\mathrm{Mpc}$; all perturbations are therefore of order the BAO scale itself, which makes the test conservative, since any acoustic response induced by kernel mismatch at these magnitudes should be detectable. The perturbation is applied to the \emph{full} ETG and LTG catalogues before the peaked-bin selection, rather than to the already-selected bin. This ensures that the perturbed redshifts propagate into both the sample selection and the overdensity map, so that galaxies crossing in and out of the window contribute to the change in the $C_{\ell}$ ratio; if the peaked-bin selection is performed first and the perturbation is applied afterwards, the perturbation has no effect on the angular overdensity $\delta(p) = n_{\mathrm{data}}(p)/\bar{n}_{\mathrm{rand}}(p) - 1$ and every configuration returns a bit-identical posterior. The random catalogues are regenerated from the perturbed data redshifts in each configuration, so that any mean shift or broadening of the kernel is absorbed self-consistently into the $n_{\mathrm{data}}/n_{\mathrm{rand}}$ ratio.

\begin{figure*}[t]
\centering
\includegraphics[width=0.98\textwidth]{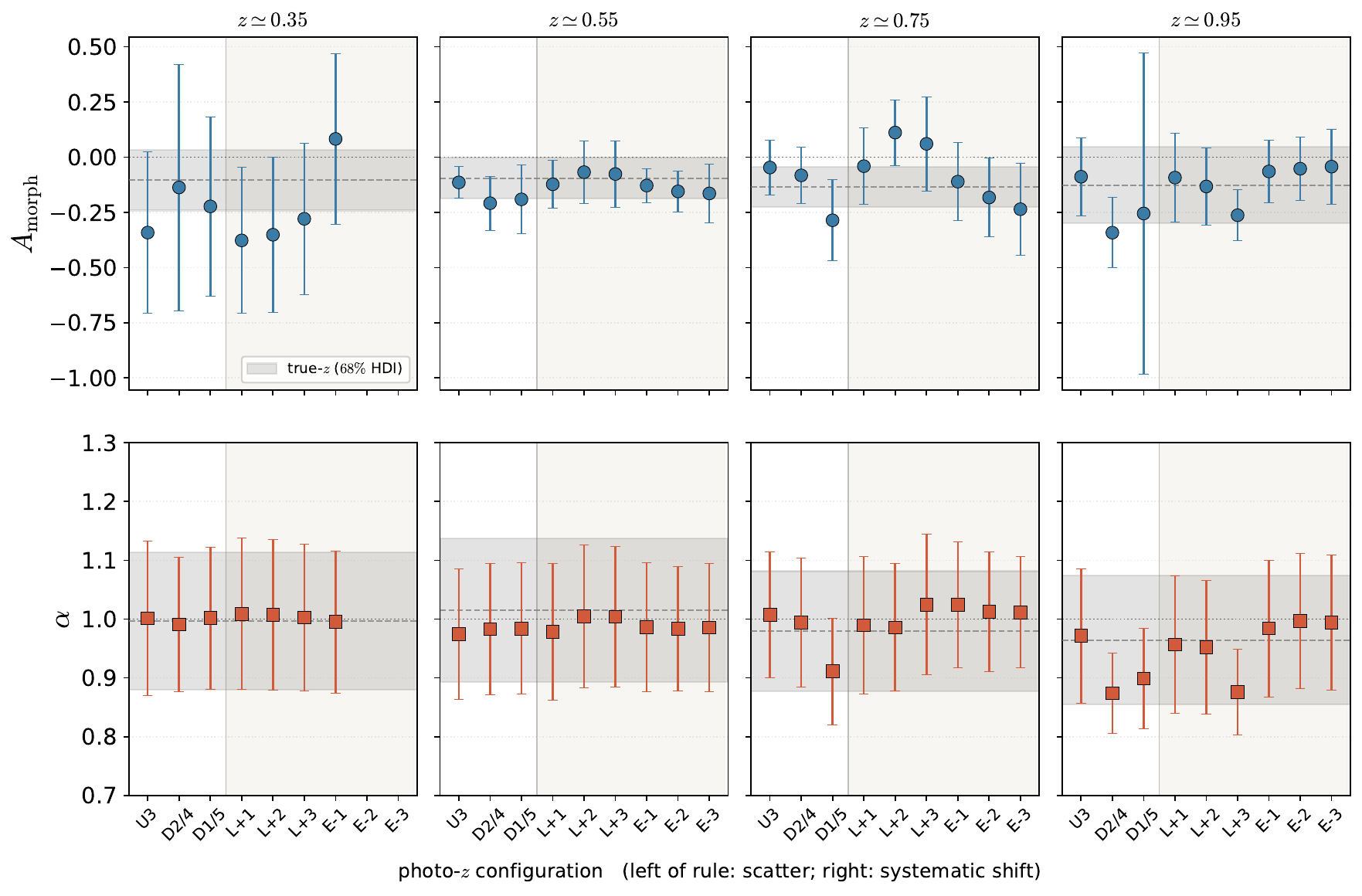}
\caption{Photo-$z$ robustness of the acoustic response estimator, tested on the Euclid Flagship~2 mock at four redshift bins. Each panel shows the absolute value of $A_{\mathrm{morph}}$ (top row) and $\alpha$ (bottom row) recovered from the CTCR pipeline after applying one of nine photo-$z$ perturbations to the morphology-selected ETG/LTG pair. The $x$-axis is shared across all panels and labels are compact: \texttt{U3} denotes uniform $3\%$ scatter, \texttt{D2/4} and \texttt{D1/5} differential scatter $2\%/4\%$ and $1\%/5\%$, \texttt{L+1} to \texttt{L+3} systematic LTG shifts of $1\%$ to $3\%$ in $(1+z)$, and \texttt{E-1} to \texttt{E-3} the corresponding ETG shifts. A vertical rule separates the scatter group (left) from the shift group (right). The grey band in each panel is the true-redshift reference and its $68\%$ HDI; the dotted line marks the null value $A_{\mathrm{morph}} = 0$ (top) or $\alpha = 1$ (bottom). All nine configurations lie on or within the reference band in every redshift bin within the statistical uncertainties, except for the two highest ETG shifts at $z = 0.35$, where the ETG sample after the perturbation falls below the minimum count threshold and the bin is excluded (see text). Blue circles: $A_{\mathrm{morph}}$; red squares: $\alpha$.}
\label{fig:photoz_per_redshift}
\end{figure*}

Figure~\ref{fig:photoz_per_redshift} summarises the outcome of the full CTCR acoustic response pipeline applied to each perturbed catalogue. The figure is organised as a $2 \times 4$ grid: one column per redshift bin ($z_{\rm peak} = 0.35, 0.55, 0.75, 0.95$), the top row showing the absolute value of the acoustic response amplitude $A_{\mathrm{morph}}$ and the bottom row the absolute value of the dilation parameter $\alpha$. The $x$-axis is shared across all eight panels and labels each configuration with a compact tag: \texttt{U3} for uniform $3\%$ scatter, \texttt{D2/4} and \texttt{D1/5} for the two differential-scatter configurations, \texttt{L+1}, \texttt{L+2}, \texttt{L+3} for the LTG shifts, and \texttt{E-1}, \texttt{E-2}, \texttt{E-3} for the ETG shifts. A thin vertical rule and a cream background distinguish the three scatter configurations on the left from the six systematic shifts on the right. In each panel, the true-redshift result is shown as a grey horizontal band (corresponding to its $68\%$ highest-density interval) and the null value is marked by a dotted line ($A_{\mathrm{morph}} = 0$ in the top row, $\alpha = 1$ in the bottom row). Three features stand out. First, the recovered $\alpha$ is consistent with unity in every configuration and every redshift bin, with a maximum deviation of $1.4\sigma$ for the differential-scatter $2\%/4\%$ and $1\%/5\%$ configurations at $z = 0.95$ and $z = 0.75$ respectively ($\alpha = 0.874 \pm 0.068$ and $\alpha = 0.911 \pm 0.090$); photo-$z$ errors of the magnitude expected in the DESI Legacy Survey therefore do not bias the BAO scale recovered by the CTCR estimator. Second, the recovered $A_{\mathrm{morph}}$ is consistent with the true-redshift reference in every configuration and every bin. At $z = 0.55$, where the ETG and LTG samples are both saturating the $N = 7 \times 10^{5}$ cap and the statistical uncertainties are smallest ($\sigma_{A} \sim 0.07 - 0.15$), the maximum deviation from the reference is $1.2\sigma$; at $z = 0.35$ the ETG sample is reduced after the perturbation and the individual uncertainties are correspondingly larger ($\sigma_{A} \sim 0.3 - 0.6$), so all configurations are consistent with the reference at better than $1\sigma$. Third, no configuration shows a systematic trend with the perturbation magnitude: the recovered values of $A_{\mathrm{morph}}$ and $\alpha$ fluctuate around the reference without monotonic dependence on $\sigma_{z}$ or $o$, which is the expected behaviour if the residual kernel mismatch is being absorbed into the smooth sigmoid baseline rather than into the acoustic template. Two entries are missing from the lowest-redshift panel of Fig.~\ref{fig:photoz_per_redshift}: the largest two ETG shifts at $z = 0.35$ leave fewer than $5{,}000$ ETG galaxies in the peaked bin after the perturbation, below the minimum count required for a stable jackknife covariance estimate, and the corresponding fits are excluded; the three remaining ETG shifts at that redshift are shown and are consistent with the reference at better than $1\sigma$, albeit with uncertainties a factor of three larger than at intermediate redshifts. The physical interpretation of these results is straightforward. When the random catalogue is regenerated from the perturbed redshifts, the effective radial kernel that enters the CTCR is re-estimated self-consistently, and any mean shift or broadening is absorbed into the ratio $n_{\mathrm{data}}/n_{\mathrm{random}}$. Any residual scale dependence is then absorbed by the sigmoid baseline of Eq.~\ref{eq:sigmoid_bias}, which is flexible enough to capture the smooth component of a displaced or broadened kernel. The acoustic template of Eq.~\ref{eq:acoustic_template} responds only to a shape difference at the BAO scale, and neither a symmetric broadening nor a mean shift of order $\pm 150~\mathrm{Mpc}$ creates one, because the smooth-baseline fit is free to adjust its transition scale and width to follow the perturbed ratio. This is the same mechanism that underlies the mild negative response seen in the mass-selected pairs of Section~\ref{subsec:euclid_validation}: the estimator responds to kernel mismatch only when the mismatch produces a residual acoustic shape, not merely a smooth shift. Two caveats apply to this test. First, the acoustic template used in the fit is fixed to the fiducial wiggle/no-wiggle difference, so the test measures the amplitude of that specific template rather than an unconstrained acoustic shape; a photometric perturbation that produced an acoustic residual with a different shape would be partly absorbed into the sigmoid baseline and would not be recovered as a nonzero $A_{\mathrm{morph}}$. Second, the test is performed on a dark-matter-only light cone with no intrinsic alignment or baryonic feedback, so the results constrain the projection and selection systematics of the CTCR but not the astrophysical systematics of the morphological split itself. Bearing these caveats in mind, the conclusion of the photo-$z$ robustness test is that the acoustic response estimator is insensitive to photo-$z$ errors of the magnitude expected in the DESI Legacy Survey, in both scatter and mean. The systematic shifts seen in the mass-selected pairs of Section~\ref{subsec:euclid_validation} are therefore not a photo-$z$ artefact: they are driven by the intrinsic difference in the shapes of the two $n(z)$ distributions, which is present even in the true-redshift mock. The DESI morphology-selected result, $A_{\mathrm{morph}} = 0.37 \pm 0.31$, is therefore not contaminated by photometric redshift systematics, and the bound on the differential acoustic response between ETGs and LTGs is robust.

\section{Summary}

We have presented a measurement of the angular BAO scale from the DESI Legacy Survey DR8 using CTCR, and we have introduced a new estimator for the differential acoustic response between two morphology-selected galaxy populations. 

By deriving the CTCR ratio starting from the galaxy overdensity field and a renormalized Eulerian bias expansion. In the matched-kernel deterministic linear-bias limit, the complete projected matter power spectrum cancels in the ratio, including both the smooth broadband and the BAO wiggles. Any residual scale dependence must therefore originate from a difference between the normalized non-linear responses of the two tracers. The measured ratio is modelled as a smooth scale-dependent baseline multiplied by a scale-dependent acoustic template which is our acoustic response estimator of Eq~\ref{eq:acoustic_response_model}. It contains two physical parameters: the amplitude \(A_{\rm morph}\) of the differential acoustic response and the BAO dilation parameter \(\alpha\).

Combining the nine bins. we obtained $\alpha^{\rm DESI}=0.9997 \pm 0.0383$ consistent with the fiducial cosmology based on $\Lambda$CDM's Planck 18 at \(0.01\sigma\) level. All values of each bins are consist with unity, and the precision of the combined constraint is  approximately 3.8 \%, compared to the Euclid F2 validation and DES Y6 angular BAO measurements \cite{DESY6}. 

The combined acoustic response amplitude is \(A_{\rm morph}^{\rm DESI} =0.369 \pm 0.307\)  consistent with the null hypothesis \(A_{\rm morph} = 0\) at the \(1.2\sigma\) level. All nine bins are individually consistent with zero within \(1\sigma\), and there is no systematic trend with
redshift. We conclude that the DESI morphology-selected ETG and LTG samples do not exhibit a detectable differential acoustic response at the current level of precision.

A comparison of the smooth-only model (8 parameters) with the full model (10 parameters) using WAIC, BIC, and AIC gives consistent results: the full model is marginally preferred in six of the nine redshift bins, with \(\Delta\)WAIC significances of at most \(2.6\sigma\), and the two models are indistinguishable in the remaining three bins. The \(\Delta\)BIC values are systematically more negative but are not used for model selection because the number of data points per bin is small. The model comparison therefore confirms that the data do not require an acoustic modulation of the CTCR ratio beyond the smooth baseline.

The CTCR ratio is not, in general, equal to the physical bias ratio, but it was useful to estimate an effective bias scale dependence \cite{Ferreira:2026iru}. The two morphological populations are imperfectly correlated, with \(r_\ell = 0.57\)--\(0.81\) across the DESI redshift range, deviating from unity at \(15.5\sigma\) to \(25.8\sigma\) significance. The true bias ratio is therefore $\frac{b_E}{b_L}=\mathcal{R}_\ell^{EL} \, r_\ell$. Correcting for imperfect correlation, the bias ratio across DESI sample is \(b_E/b_L \sim 1.2\)--\(1.7\), consistent with the linear bias
measurements of \citet{Ferreira:2026iru}. This can be explained by the different halo occupation statistics of the two morphological types, consistent with the halo assembly bias framework.

We validated the acoustic response estimator on the Euclid F2 mock across six redshift bins and four tracer pairs. Through a pre-defined morphology gives \(A_{\rm morph} \simeq 0\) and \(\alpha \simeq 1\) in every bin, with uncertainties of order \(0.10\). Through mass selected pairs of compatible definitions of ETG2 and LTG shows a mild negative response and a 1$\sigma$ downward shift in $\alpha$ in intermediate bins. The LRG2/LTG pair has the largest scatter, as expected from its smaller sample size. No tracer pair shows a statistically significant positive response. The Euclid validation
therefore confirms that the estimator is sensitive to the quality of the tracer matching, returning a null result when the tracers are drawn from the same underlying population with matched redshift distributions, and correctly flagging a systematic when they are not.

We tested the estimator's sensitivity to photo-\(z\) errors on the Euclid Flagship 2 mock at four redshifts, applying Gaussian scatter up
to \(\sigma_z = 0.05\) and systematic peak shifts of one tracer by up to \(3\%\) in \((1+z)\). All configurations leave \(A_{\rm morph}\)
and \(\alpha\) unchanged within the statistical uncertainties, with maximum shifts an order of magnitude below the DESI errors. The
estimator is therefore robust to photo-\(z\) systematics of the magnitude expected in the DESI Legacy Survey, and the
morphology-selected bound on the differential acoustic response is not contaminated.

The CTCR acoustic response estimator provides a new probe of morphology-dependent clustering in photometric surveys. The DESI
Legacy Survey DR8 results are consistent with \(\Lambda\)CDM: the BAO scale is recovered without bias, and no differential acoustic response is detected between ETGs and LTGs. Future wide-area imaging surveys with precise morphological information, including \textit{Euclid}, the Nancy Grace Roman Space Telescope, the CSST, and LSST, will provide larger samples and improved redshift precision for extending this analysis to finer morphological classes and higher redshift. 
\begin{acknowledgments}

PSF and RC acknowledge support from the National Key Research and Development Program of China and the Zhejiang Provincial Top-Level Research Support Program.  CAPB acknowledges financial support from the CNPq grant 306630/2025-7. This research was carried out using the SilkRiver Supercomputer at Zhejiang University.

This work has made use of CosmoHub, developed by PIC (maintained by IFAE and CIEMAT) in collaboration with ICE-CSIC. It received funding from the Spanish government (grant EQC2021-007479-P funded by MCIN/AEI/10.13039/501100011033), the EU NextGeneration/PRTR (PRTR-C17.I1), and the Generalitat de Catalunya.

The Legacy Surveys consist of three individual and complementary projects: the Dark Energy Camera Legacy Survey (DECaLS; Proposal ID \#2014B-0404; PIs: David Schlegel and Arjun Dey), the Beijing-Arizona Sky Survey (BASS; NOAO Prop. ID \#2015A-0801; PIs: Zhou Xu and Xiaohui Fan), and the Mayall z-band Legacy Survey (MzLS; Prop. ID \#2016A-0453; PI: Arjun Dey). DECaLS, BASS and MzLS together include data obtained, respectively, at the Blanco telescope, Cerro Tololo Inter-American Observatory, NSF’s NOIRLab; the Bok telescope, Steward Observatory, University of Arizona; and the Mayall telescope, Kitt Peak National Observatory, NOIRLab. Pipeline processing and analyses of the data were supported by NOIRLab and the Lawrence Berkeley National Laboratory (LBNL). The Legacy Surveys project is honored to be permitted to conduct astronomical research on Iolkam Du’ag (Kitt Peak), a mountain with particular significance to the Tohono O’odham Nation.

NOIRLab is operated by the Association of Universities for Research in Astronomy (AURA) under a cooperative agreement with the National Science Foundation. LBNL is managed by the Regents of the University of California under contract to the U.S. Department of Energy.

This project used data obtained with the Dark Energy Camera (DECam), which was constructed by the Dark Energy Survey (DES) collaboration. Funding for the DES Projects has been provided by the U.S. Department of Energy, the U.S. National Science Foundation, the Ministry of Science and Education of Spain, the Science and Technology Facilities Council of the United Kingdom, the Higher Education Funding Council for England, the National Center for Supercomputing Applications at the University of Illinois at Urbana-Champaign, the Kavli Institute of Cosmological Physics at the University of Chicago, Center for Cosmology and Astro-Particle Physics at the Ohio State University, the Mitchell Institute for Fundamental Physics and Astronomy at Texas A\&M University, Financiadora de Estudos e Projetos, Fundacao Carlos Chagas Filho de Amparo, Financiadora de Estudos e Projetos, Fundacao Carlos Chagas Filho de Amparo a Pesquisa do Estado do Rio de Janeiro, Conselho Nacional de Desenvolvimento Cientifico e Tecnologico and the Ministerio da Ciencia, Tecnologia e Inovacao, the Deutsche Forschungsgemeinschaft and the Collaborating Institutions in the Dark Energy Survey. The Collaborating Institutions are Argonne National Laboratory, the University of California at Santa Cruz, the University of Cambridge, Centro de Investigaciones Energeticas, Medioambientales y Tecnologicas-Madrid, the University of Chicago, University College London, the DES-Brazil Consortium, the University of Edinburgh, the Eidgenossische Technische Hochschule (ETH) Zurich, Fermi National Accelerator Laboratory, the University of Illinois at Urbana-Champaign, the Institut de Ciencies de l’Espai (IEEC/CSIC), the Institut de Fisica d’Altes Energies, Lawrence Berkeley National Laboratory, the Ludwig Maximilians Universitat Munchen and the associated Excellence Cluster Universe, the University of Michigan, NSF’s NOIRLab, the University of Nottingham, the Ohio State University, the University of Pennsylvania, the University of Portsmouth, SLAC National Accelerator Laboratory, Stanford University, the University of Sussex, and Texas A\&M University.

BASS is a key project of the Telescope Access Program (TAP), which has been funded by the National Astronomical Observatories of China, the Chinese Academy of Sciences (the Strategic Priority Research Program “The Emergence of Cosmological Structures” Grant \# XDB09000000), and the Special Fund for Astronomy from the Ministry of Finance. The BASS is also supported by the External Cooperation Program of Chinese Academy of Sciences (Grant \# 114A11KYSB20160057), and Chinese National Natural Science Foundation (Grant \# 12120101003, \# 11433005).

The Legacy Survey team makes use of data products from the Near-Earth Object Wide-field Infrared Survey Explorer (NEOWISE), which is a project of the Jet Propulsion Laboratory/California Institute of Technology. NEOWISE is funded by the National Aeronautics and Space Administration.

The Legacy Surveys imaging of the DESI footprint is supported by the Director, Office of Science, Office of High Energy Physics of the U.S. Department of Energy under Contract No. DE-AC02-05CH1123, by the National Energy Research Scientific Computing Center, a DOE Office of Science User Facility under the same contract; and by the U.S. National Science Foundation, Division of Astronomical Sciences under Contract No. AST-0950945 to NOAO.

The Photometric Redshifts for the Legacy Surveys (PRLS) catalog used in this paper was produced thanks to funding from the U.S. Department of Energy Office of Science, Office of High Energy Physics via grant DE-SC0007914.

\end{acknowledgments}

\appendix

\newpage
\bibliographystyle{apsrev4-2}
\bibliography{ref}
\end{document}